\documentclass[sn-mathphys,Numbered]{sn-jnl}% Math and Physical Sciences Reference Style

\usepackage{graphicx}%
\usepackage{multirow}%
\usepackage{amsmath,amssymb,amsfonts}%
\usepackage{amsthm}%
\usepackage{mathrsfs}%
\usepackage[title]{appendix}%
\usepackage{xcolor}%
\usepackage{textcomp}%
\usepackage{manyfoot}%
\usepackage{booktabs}%
\usepackage{algorithm}%
\usepackage{algorithmicx}%
\usepackage{algpseudocode}%
\usepackage{listings}%https://cn.overleaf.com/project
\usepackage{color}
\usepackage{booktabs} %需要加载宏包{booktabs}
\usepackage{mathrsfs}
\usepackage{siunitx}[=v2] %国际单位制宏包
\begin{document}

\title{Anomalous large-scale collective motion in granular Brownian vibrators}

\author[1]{Yangrui Chen}\email{chenyangrui66@gmail.com}
\author*[1,2]{Jie Zhang}\email{jiezhang2012@sjtu.edu.cn}

\affil*[1]{School of Physics and Astronomy, Shanghai Jiao Tong University, Shanghai 200240, China}
\affil*[2]{Institute of Natural Sciences, Shanghai Jiao Tong University, Shanghai 200240, China}

\abstract{
Using Brownian vibrators, we conducted a study on the structures and dynamics of quasi-2d granular materials with packing fractions ($\phi$) ranging from 0.111 to 0.832. Our observations revealed a remarkable large-scale collective motion in hard granular disk systems, encompassing four distinct phases: granular fluid, collective fluid, poly-crystal, and crystal. The collective motion emerge at $\phi=$0.317, coinciding with a peak in local density fluctuations. However, this collective motion ceased to exist at $\phi=$0.713 when the system transitioned into a crystalline state.
While the poly-crystal and crystal phases exhibited similarities to equilibrium hard disks, the first two phases differed significantly from the equilibrium systems and previous experiments involving uniformly driven spheres. This disparity suggests that the collective motion arises from a competition controlled by volume fraction. Specifically, it involves an active force and an effective attractive interaction resulting from inelastic particle collisions. Remarkably, these findings align with recent theoretical research on the flocking motion of spherical active particles without alignment mechanisms.}


\maketitle

\section{Introduction}\label{sec1}

Collective motion, characterized by the coordinated behavior of microscopic components on a large scale in both space and time, is a pervasive phenomenon observed in various systems, including soft matter and active matter \cite{Nagel-RevModPhys.89.025002,chaikin_lubensky_1995,shankar2022topological,shaebani2020computational}.
In the realm of active matter, the mechanisms driving collective motion have been extensively studied. For instance, the collective behavior of bird flocks can be effectively modeled as a system of ``flying spins," where the alignment of neighboring individuals' moving directions plays a crucial role \cite{vicsek1995novel, Toner-Tu-PhysRevLett.75.4326, Toner-Tu-PhysRevE.58.4828}. Similarly, rod-like particles can be described as active nematics, incorporating factors such as local alignment and volume exclusion to explain their collective motion \cite{doostmohammadi2018active}. Notably, when particles possess intrinsic spinning motion, intriguing topologically protected edge modes can arise due to nonreciprocal interactions \cite{banerjee2017odd}.
These examples highlight the profound understanding gained in elucidating the mechanisms governing collective motion in active matter systems.\par
Collective motion in granular materials is often associated with phenomena such as jamming, particle polarity, and particle shape. In densely packed granular systems subjected to quasi-static shear, the behavior of contacts and contact forces plays a crucial role in floppy modes\cite{liu2010jamming, van2009jamming}, plastic deformation\cite{maloney2006amorphous,wang2020connecting}, and the formation of turbulent-like vortices\cite{Farhang-PhysRevLett.89.064302, sun2022turbulent}. When granular materials are subjected to vibration, inelastic collisions become prominent, and the polarity and shape of the particles become significant factors. Systems composed of self-propelling polar particles\cite{deseigne2010collective, deseigne2012vibrated, kudrolli2008swarming} and self-spinning disks\cite{scholz2018rotating,liu2020oscillating} exhibit collective behaviors, blurring the line between granular materials and active matter. Additionally, the behavior of rod-like particles can be effectively described using the framework of active nematics \cite{kumar2014flocking, kudrolli2008swarming, narayan2010phase,kumar2011symmetry, doostmohammadi2018active}. 
Simulations have shown that the alignment interaction plays a critical role in the flocking behavior of self-propelling purely repulsive hard disks \cite{fily2012athermal}. In the absence of this alignment interaction, only motility-induced phase separation is observed \cite{redner2013structure, buttinoni2013dynamical}. Previous experiments with purely repulsive hard spheres or disks, lacking preferred translational or rotational directions, under uniform driving in a 2d system, have not exhibited any collective motion \cite{olafsen1998clustering, Shattuck-PhysRevLett.98.188301, losert1999velocity, tatsumi2009experimental, melby2005dynamics}.\par
However, recent theoretical work by Caprini et al. \cite{caprini2023flocking} challenges the prevailing notion that alignment interactions are a prerequisite for the emergence of flocking behavior. Their study predicts that flocking can occur even in the absence of explicit alignment interactions. Instead, the transition between disordered and flocking phases is primarily determined by the interplay between the active forces and the attractive forces arising from neighboring particles. Since inelastic collision is analogous to an effective attraction, conducting experiments with vibrated granular materials may provide a promising avenue for experimentally verifying this prediction, which remains unverified to date.\par
In this letter, we present a systematic investigation of the structures and collective dynamics in quasi-2d experiments using Brownian vibrators. These vibrators consist of disks with alternating inclined legs along the rim. By exploring a wide range of packing fractions ($\phi$), we uncover a rich variety of structures and dynamics arising from inelastic particle collisions. Specifically, we observe the emergence of distinct phases, namely the granular fluid, collective fluid, poly-crystal, and crystal.
Interestingly, while the poly-crystal and crystal phases exhibit striking similarities to equilibrium hard disks\cite{Mitus-PRE-1997, alder1957phase, Zollweg-PRB, Lee-PRB, Weber-EPL, Alonso-PRL}, the first two phases differ significantly from both the equilibrium systems and previous quasi-2d experiments involving uniformly driven spheres \cite{olafsen1998clustering, Shattuck-PhysRevLett.98.188301, Olafsen-Urbach-2005, Pacheco-Vazquez-2009, Aranson-PRL-2000, Howell-PRE-2001, Oyarte-PRE-2013, Rivas-2011a, Rivas-2011b, Rivas-2012, Roeller-PRL-2011, Nahmad-PRE, Neel-PRE-2004, Clewett-2012, Luu-2013}. Remarkably, our experiments demonstrate the occurrence of large-scale collective particle motion within the range of $0.317\leq \phi < 0.713$, which aligns remarkably well with the theoretical prediction made by Caprini et al.\cite{caprini2023flocking}. Overall, our investigation reveals that granular materials subjected to uniform stochastic driving exhibit weak cohesion, intricate internal structures, and dynamic behaviors. Furthermore, our findings highlight the possibility of large-scale collective motion in purely repulsive hard-disk systems, broadening our understanding of the collective dynamics in granular materials.\par

\section{Results}\label{sec2}
Fig. \ref{fig1} provides snapshots of particle configurations representing the (a)granular fluid, (b)collective fluid, (c)poly-crystal phases, and (d)crystal phases.
In the pre-experiment, each individual particle exhibits Brownian-like behavior, characterized by uncorrelated translation and rotation with Gaussian distributions centered around zero means, in the absence of particle-particle collisions\cite{chen2022high}.

\subsection{Dynamics.} 
\begin{figure}[htpb]
\centering
\includegraphics[width=1\textwidth]{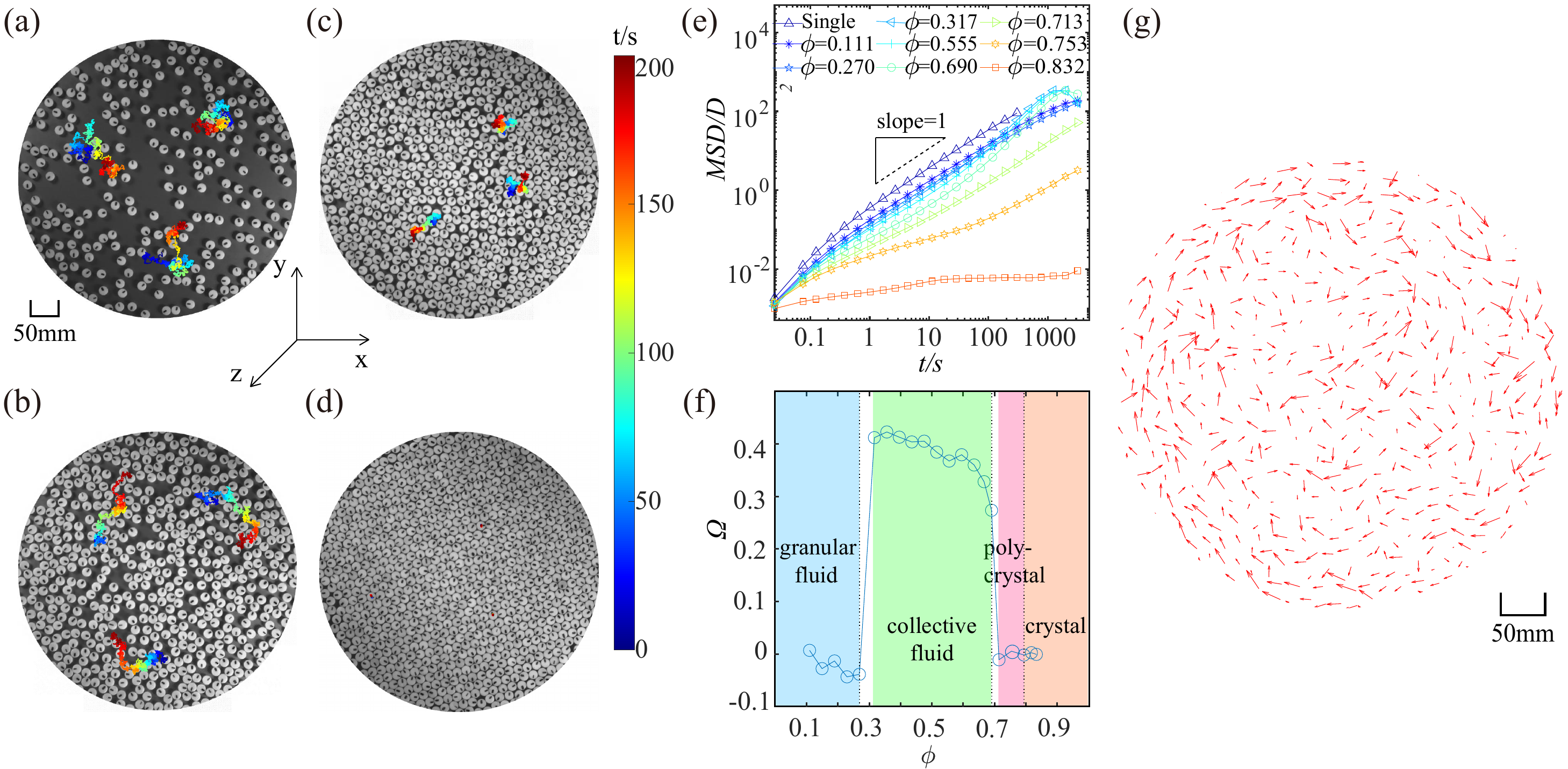}
\caption{(a-d) Snapshots of particle configurations at $\phi=0.270,0.555,0.713,0.832$, respectively. The color bar represents the selected particles’ positions in 0-200s. (e) MSD of the particle’s translational motion. (f) Granular fluid ($\phi\le 0.270$), collective fluid ($0.317 \le \phi\le\ 0.690$), poly-crystal ($0.713 \le \phi\le\ 0.753$), and crystal ($\phi \ge 0.793$). $\Omega$(blue circle) is the average curl of the particle displacement field within $\Delta t=100s$. (g) Particle displacement field at $\phi=0.555$. The length of the arrows is scaled to half of the displacement to enhance visibility. }
\label{fig1} 
\end{figure}

In Figs.~\ref{fig1} (a-d), we draw three particles' trajectories to illustrate different dynamics. At $\phi=0.270$, the three particles perform random walks at all times in Fig.~\ref{fig1} (a). At $\phi=0.555$, particles move randomly for $\Delta t<10s$, and however, for $\Delta t>100s$ they all traverse a clockwise arc around the center of the plane in Fig.~\ref{fig1}(b), suggesting a collective motion. At $\phi=0.713$, particles diffuse around slowly in Fig.~\ref{fig1} (c). At $\phi=0.832$, the particles oscillate near the lattice like a crystal in Fig.~\ref{fig1} (d). \par

The translational mean square displacements (MSD) are illustrated in Fig.~\ref{fig1} (e). For $\phi \le 0.270$, the particles exhibit superdiffusive behavior for $t\lesssim0.2s$ before transitioning to normal diffusion, indicating that a particle persistence time of $ \tau_p= 0.2\pm0.05s$ at $\phi \le 0.270$, as shown in Fig. ~\ref{sfig2} of Appendix A.
The slope decreases slightly below one for $t\gtrsim100s$.  When $\phi=0.317,0.555,0.690$ in the collective fluid phase,
particles are superdiffusive for $t\lesssim0.1 s$ before turning subdiffusive for $0.1s\lesssim t\lesssim 20s$.
However, when $t\gtrsim20s$, the slope is above one, showing superdiffusion due to the large-scale collective motion. 
Moreover, the MSD peaks around 2000s correspond to a half period of the collective rotation of the outermost particles. These particles contribute the most to the MSD and travel from one side of the system to the other through an approximately semicircular arc path, reaching the system size scale (about 30 particle diameters after the removal of the boundary particles). After this point, the MSD begins to fall.
The MSD of $\phi=0.713$ is divided into three parts: 
the superdiffusion for $t \lesssim 0.05s$, the subdiffusion for $0.05 \lesssim t \lesssim 30s$, and the diffusion for $t\gtrsim 30s$ due to particles at grain boundaries (See Figs. ~\ref{sfig4}-~\ref{sfig5} of Appendix B and C). As $\phi$ increases to $\phi=0.753$, the MSD slope gradually decreases and eventually down to 0 till $\phi=0.793$, beyond which e.g., at $\phi=0.832$, only a single crystal exists. We characterize the large-scale collective motion using the nonzero vorticity $\Omega$ -- the average curl of the particle displacement field in Fig.~\ref{fig1}(f), whose computation details is in the Appendix D and E. Fig.~\ref{fig1}(g) depict the displacement fields within $\Delta t=100s$  for structures identical to Fig.~\ref{fig1}(b).
The crystallization above $\phi=0.713$ in Fig.~\ref{fig1}(f) is similar to the early experiment\cite{Shattuck-PhysRevLett.98.188301}, where spheres were sandwiched and vertically vibrated between two horizontal plates, and when $0.652< \phi< 0.719$,  their system shows subdiffusive, caging-type behaviors on MSD at intermediate time scales, similar to the curve of $\phi=0.690$ in Fig.~\ref{fig1}(f). However, within $0.652< \phi< 0.719$, no large-scale motions are observed \cite{Shattuck-PhysRevLett.98.188301}; a so-called “isotropic fluid phase” was observed for $\phi<0.652$ \cite{ Shattuck-PhysRevLett.98.188301, shattuck-PRL-crystal}. Moreover, our system is locally more ordered as shown in Figs. ~\ref{sfig5}-~\ref{sfig6}  of the Appendix C.
Interestingly, $\phi=0.713$ of poly-crystal in Fig.\ref{fig1}(f) is nearly identical to the melting-transition point $\phi_s=0.716$ predicted for the equilibrium hard disks \cite{Mitus-PRE-1997}. Furthermore, $\phi=0.690$ of collective fluid in Fig.\ref{fig1}(f) is identical to the value of the pure fluid $\phi_f$ \cite{Mitus-PRE-1997}.  Note that the precise values of $\phi_s$ and $\phi_f$ may vary slightly\cite{Mitus-PRE-1997, alder1957phase, Zollweg-PRB, Lee-PRB, Weber-EPL, Alonso-PRL}. We do not have data points within $0.690<\phi<0.713$ due to the discrete increment of $\phi$. However, particle configurations at $\phi=0.690$ and $0.713$ show different symmetries in Fig. ~\ref{sfig7} of Appendix C.

\begin{figure}[htpb]
\centering
\includegraphics[width=0.9\textwidth]{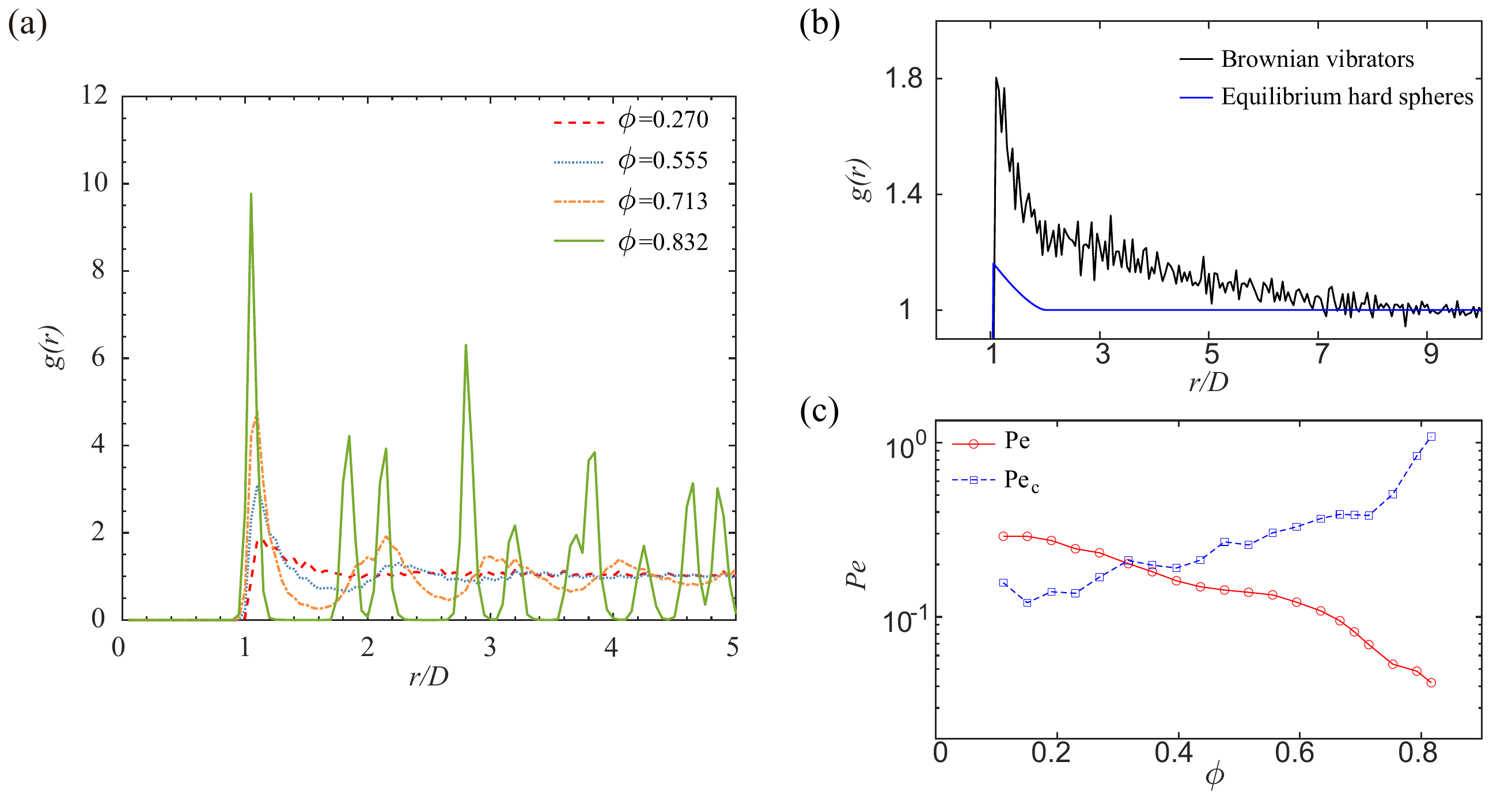}
\caption{\label{fig2} (a) Pair correlation functions $g(r)$. (b) $g(r)$ at $\phi=0.111$. (c) The P\'{e}clet number Pe of active particles and the critical value $Pe_c$, which intersect around $\phi=0.317$.}
\end{figure}

\subsection{Structure.} 
Fig.~\ref{fig2} (a) plots pair correlation functions $g(r)$ of different phases, whose first peak corresponds to the maximum possible center-to-center distance $d_1$ between neighboring particles. At $\phi=0.832$, the sharp peaks at $\sqrt{3}d_1$, $2d_1$ and $ \sqrt{7}d_1$ indicate an almost perfect crystal. At $\phi =0.713$, the system forms a poly-crystal, showing a slightly larger value of $d_1$ on $g(r)$ than that of $\phi=0.832$. Moreover, the peak at $\sqrt{3}d_1$ symbolizes the triangular lattice. Still, it is almost buried within an extensive shoulder of the peak around $2d_1$ due to grain boundaries.  At $\phi =0.555$, the system forms a collective fluid, where the first peak shifts further right than that of $\phi =0.713$. The crystalline feature is washed out as the double peaks near $\sqrt{3}d_1\sim 2d_1$ disappear entirely and are replaced with a broad single peak, indicating liquid-like structures. Moreover, there are no visible peaks after the second peak in contrast to the poly-crystal. \par

Intriguingly, in the $g(r)$ for $\phi=0.270$, the second peak disappears, while the first peak remains intact. This observation suggests that neighboring particles have a tendency to stay in close proximity, forming pairs. In Fig.~\ref{fig2}(b), we observe that at $\phi=0.111$, the peak of $g(r)$ is notably higher than that of equilibrium hard disks\cite{chae1969radial}. This disparity arises due to the dissipative nature of inelastic collisions, which allows particles to approach each other more closely than they would in equilibrium. For a detailed comparison of $g(r=D)$ versus $\phi$ with equilibrium hard disks and early experiments \cite{shattuck-PRL-crystal}, please refer to Fig. ~\ref{sfig1} of Appendix A. \par

Assuming that the inelastic collisions can be effectively represented by an attractive potential, according to Boltzmann's law, the $g(r) = \exp(-V(r)/(kT)) \cdot g_{\text{eq}}(r)$. The equilibrium $g_{\text{eq}}(r)$ for the purely repulsive hard spheres is obtained from the BGY theory in Ref.\cite{chae1969radial}. The interparticle force $F_i= -\frac{d(V(r))}{dr}$ and its maximum: \par
\begin{align}
F_{max} = \frac{1}{2}m v_T^2 \left[\frac{d\ln\left(\frac{g(r)}{g_{eq}(r)}\right)}{dr}\right]_{max}. 
\end{align}
Here the thermal velocity $v_T=1.4\pm 0.1D/s$ is weakly affected by $\phi$ \cite{chen2022high}. From the theoretical work by Caprini et al.\cite{caprini2023flocking}, the dynamics of active particles reads:\par
\begin{align}
m\Dot{\mathbf{v_i}} =-\gamma \mathbf{v_i}+\mathbf{F_i}+\gamma_0 v_0\mathbf{n_i}+\sqrt{2\Lambda\gamma}\mathbf{\eta_i}, 
\end{align}
\begin{align}
\Dot{\mathbf{\theta_i}} =\sqrt{2D_r}\mathbf{\Theta_i}  
\end{align}
Here $\mathbf{\eta_i}$ and $\mathbf{\Theta_i}$ are white noises with zero average and unit variance, $\Lambda$ is a constant, and $m$ refers to the particle mass. The damping coefficient $\gamma$ describes the dissipation of kinetic energy. 
Unlike the simulation where the active force is given as $\gamma v_0$ \cite{caprini2023flocking}, in our experiment, the magnitude of active force $\gamma_0 v_0$ originates from the friction force at the bottom due to the unique structure of the inclined legs of the particles (refer to the Methods section).
The diffusion coefficient $D_r$ of the particle velocity direction, the active velocity $v_0$, and the friction factor $\gamma_0 \equiv m/\tau_I$ can all be determined experimentally in Appendix A. $\tau_I$ is the inertia time. 
The P\'{e}clet number $Pe=v_0/(D_rD)$ is viewed as the ratio between persistence length and particle size, quantifying the distance over which a particle maintains directional persistence.
Comparing the maximum equivalent attractive force and the active force, the critical value $Pe_c=F_{max}/(\gamma_0 v_0) \cdot Pe$ is positively correlated with the collision frequency through $F_{max}$. As shown in Fig.~\ref{fig2}(c), $Pe$ decreases while $Pe_c$ increases with $\phi$ since the persistence time $\tau_p=1/D_r$ decreases with $\phi$ meanwhile the collision frequency increases with $\phi$. At $\phi\le 0.270$, Pe is greater than $Pe_c$, indicating that the active forces dominate over the attractive forces, leading to the random motion. At $0.317 \le \phi\le 0.690$, where the maximum attractive force exceeds the active forces, long-time scale collective motion emerges. For specific details, please refer to Appendix A. This finding aligns with the theoretical work by Caprini et al.\cite{caprini2023flocking}, where the emergence of collective behavior occurs when the maximum force exerted by neighboring particles surpasses the active forces of the individual particles. Due to the periodic boundary conditions in the simulation, the collective behavior appears as flocking. However, in our quasi-2d system with fixed boundaries, the collective behavior manifests as global collective rotation around the center of the system.\par

\begin{figure}[htpb]
\centering
\includegraphics[width=0.9\textwidth]{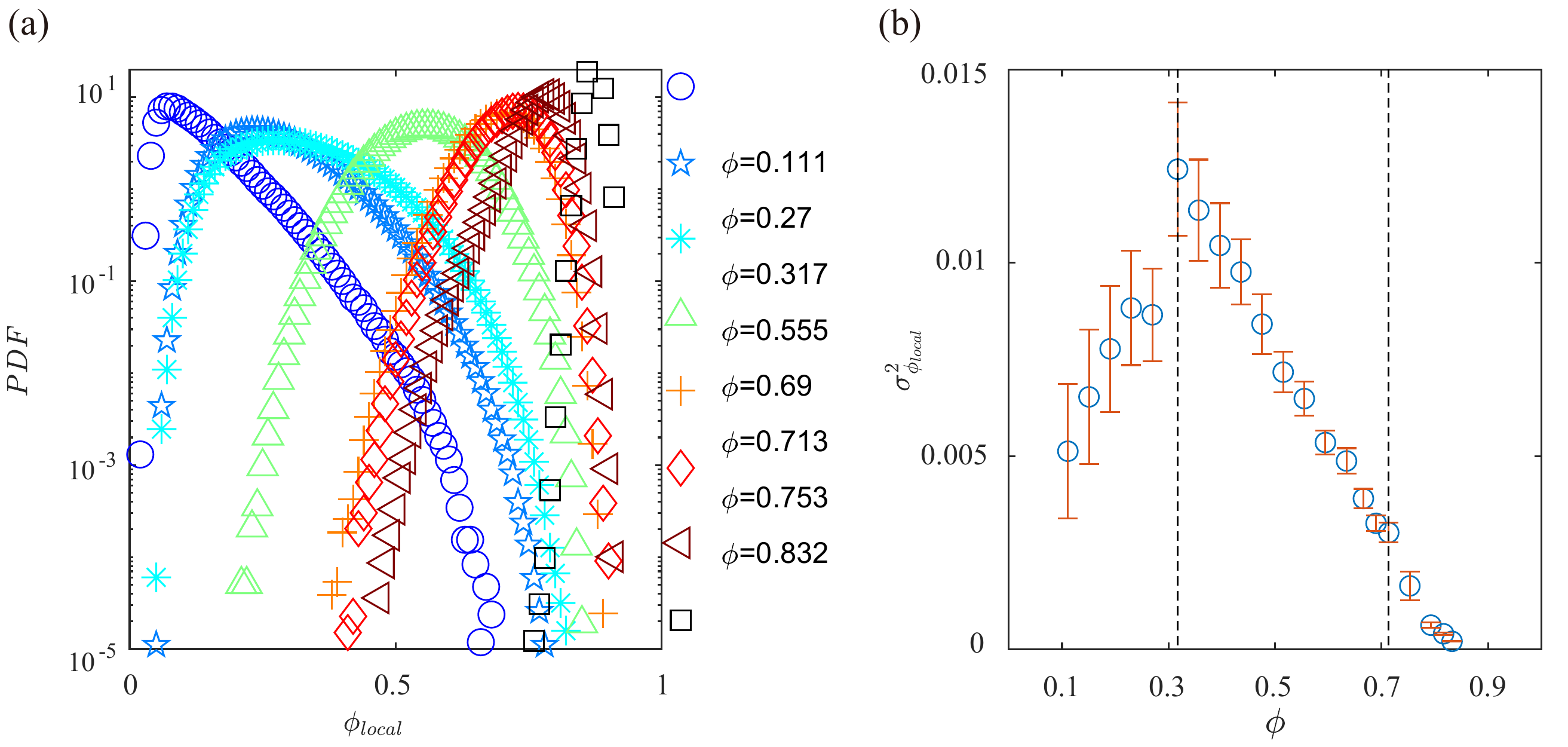}
\caption{\label{fig3} (a) The local packing fraction distribution. (b) The fluctuation of $\phi_{local}$. }
\end{figure}

The local packing fraction of disks with diameter D is determined by the formula $\phi_{local}=\frac{\pi D^2}{4 S_v}$, where $S_v$ denotes the area of the corresponding Voronoi polygons as shown in Fig.~\ref{sfig4} of Appendix B. The distribution of $\phi_{local}$ does not exhibit clear bimodal peaks at any given $\phi$. This observation suggests that in a spatially homogeneous driven quasi-2d granular system, the typical clustering or solid-liquid phase separation observed in previous boundary-driven systems is not present\cite{kudrolli1997cluster}. These findings are consistent with previous experiments involving spatially homogeneous driving conditions\cite{scholz2017velocity}.
Here the energy is constantly injected and dissipated, resulting in a driven steady state where neighboring particles do not stay in close contact due to stochastic driving separating them, in contrast to free cooling systems \cite{goldhirsch1993clustering, luding1999cluster}. Likewise, in the phase-separated systems\cite{esipov1997granular,olafsen1998clustering, Roeller-PRL-2011, Nahmad-PRE, Neel-PRE-2004, Clewett-2012, Luu-2013}, segregated cluster particles are in close contact, forming a highly compact solid due to the velocity-dependent energy injection rate.
Fig.~\ref{fig2} provides microscopic physical evidence that dry granular materials subjected to stochastic driving form weak-cohesion fluids at sufficiently low densities, consistent with the early experiments of the spinodal phase separation \cite{Clewett-2012} and the capillary-like interface fluctuations \cite{ Luu-2013} in cohesionless granular systems.
\par
A noteworthy finding is the consistent increase in the fluctuations of $\phi_{local}$ with $\phi$ in the granular fluid phase, as illustrated in Fig.~\ref{fig3}(b). This increase culminates in a peak at the onset of collective behavior, precisely at $\phi=0.317$. Subsequently, in the collective fluid phase, a monotonic decrease in the fluctuations of $\phi_{local}$ is observed. This behavior aligns precisely with the phase transition point corresponding to the emergence of long timescale collective behavior observed in the dynamics of the system.\par

\begin{figure}[htpb]
\centering
\includegraphics[width=0.9\textwidth]{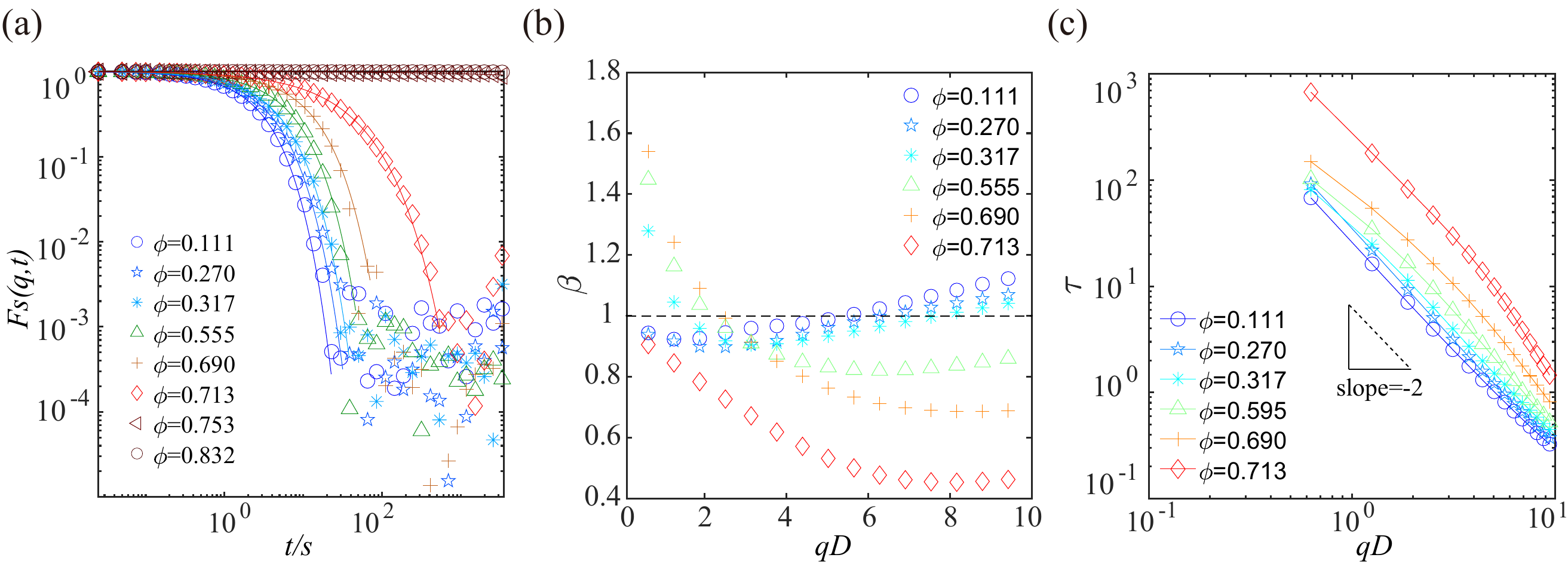}
\caption{\label{fig4} (a) Self intermediate scattering functions $F_s(q,t)$ with $q=\pi/D$. The solid lines are fitting $F_s(q,t)= exp(-(t/\tau)^{\beta})$. (b-c) The $\beta$ and $\tau$ versus q.}
\end{figure}

The intermediate scattering function $F_s(q,t)$ characterizes structural relaxation for a given wavevector $\vec q$:\par
\begin{align}
    F_s(q,t) &= \frac{1}{N}\Sigma_{j=1}\langle e^{-i\vec q\cdot (\vec r_j(t)- \vec r_j(0))} \rangle  \nonumber \\
            &=\frac{1}{N}\Sigma_{j=1}e^{-\frac{1}{4}q^2\langle (\vec r_j(t)- \vec r_j(0))^2 \rangle},
\end{align}
where $r_j(t)$ refers to the trajectory of the particle $j$ at time $t$, $N$ is the total particle number, and the average $\langle\rangle$ is over all possible initial configurations $\vec r_j(0)$. The second equality assumes Gaussian $\vec r_j(t)- \vec r_j(0)$ \cite{binder2011glassy}, approximately valid, as shown in Figs. ~\ref{sfig8}-~\ref{sfig10} in the Appendix D. 
\par

Fig.~\ref{fig4}(a) shows the $F_s(q,t)$ with $q=\pi/D$, corresponding to the scale of $2D$. $F_s(q,t)$ decay monotonically with an increasing relaxation time with $\phi$, except at $\phi=0.832$. We fit $F_s(q,t)$ with an stretched exponential $F_s(q,t)= exp(-(t/\tau)^{\beta})$ and plot $\beta$ and $\tau$ for different $q$ in Figs.~\ref{fig4}(b-c), respectively. There are three branches of $\beta(q)$ in Fig.~\ref{fig4}(b), corresponding to the granular fluid, collective fluid, and poly-crystal. 
When $\phi=0.111$ and $0.270$, $\beta <1$ for small $q$ and increase above one at $q\approx 2\pi/D$. The weak subdiffusion at large scales is consistent with Fig.~\ref{fig1}(f), reflecting the effect of finite boundaries. 
The collective fluid shows more complex $\phi$ dependent behaviors of $\beta(q)$.  For $q<2/D$, corresponding to the scales larger than $\pi D$, $\beta(q)>1$ for all $\phi$, showing characteristics of the large-scale collective motions.  For $q\ge2/D$, when $\phi=0.317$, $\beta$ stays slightly below one within $2/D<q<2\pi/D$ and then goes above one for $q>2\pi/D$, similar to the curves of the granular fluid. When $\phi=0.555$ and $0.690$, $\beta$ remain below one for $q>2\pi/D$ and reach around 0.7 for $\phi=0.690$, confirming subdiffusive glassy behaviors shown in Fig.~\ref{fig1}(f). 
When $\phi=0.713$ and $0.753$, $\beta(q)$ starts near one and decreases significantly with $q$, consistent with Fig.~\ref{fig1}(f), verifying highly constrained grain-boundary particles.
According to Eq.(4) and $F_s(q,t)=exp(-(t/\tau)^{\beta})$, $\tau(q)\propto q^{\kappa(q)},\kappa(q)=-\frac{2}{\beta(q)}$, consistent with Fig.~\ref{fig4}(c). For example, when $\phi=0.111$ and $0.270$, $\kappa(q)\approx-2$ with a weak dependence on $q$, whereas when $\phi=0.713$, $\kappa(q)\approx-2$ for small $q$ and approach $-4.4$ at high $q$ in Fig.~\ref{fig4} (c).
\par

\section{Methods}\label{sec11}

\begin{figure}[htpb]
\centering
\includegraphics[width=0.9\textwidth]{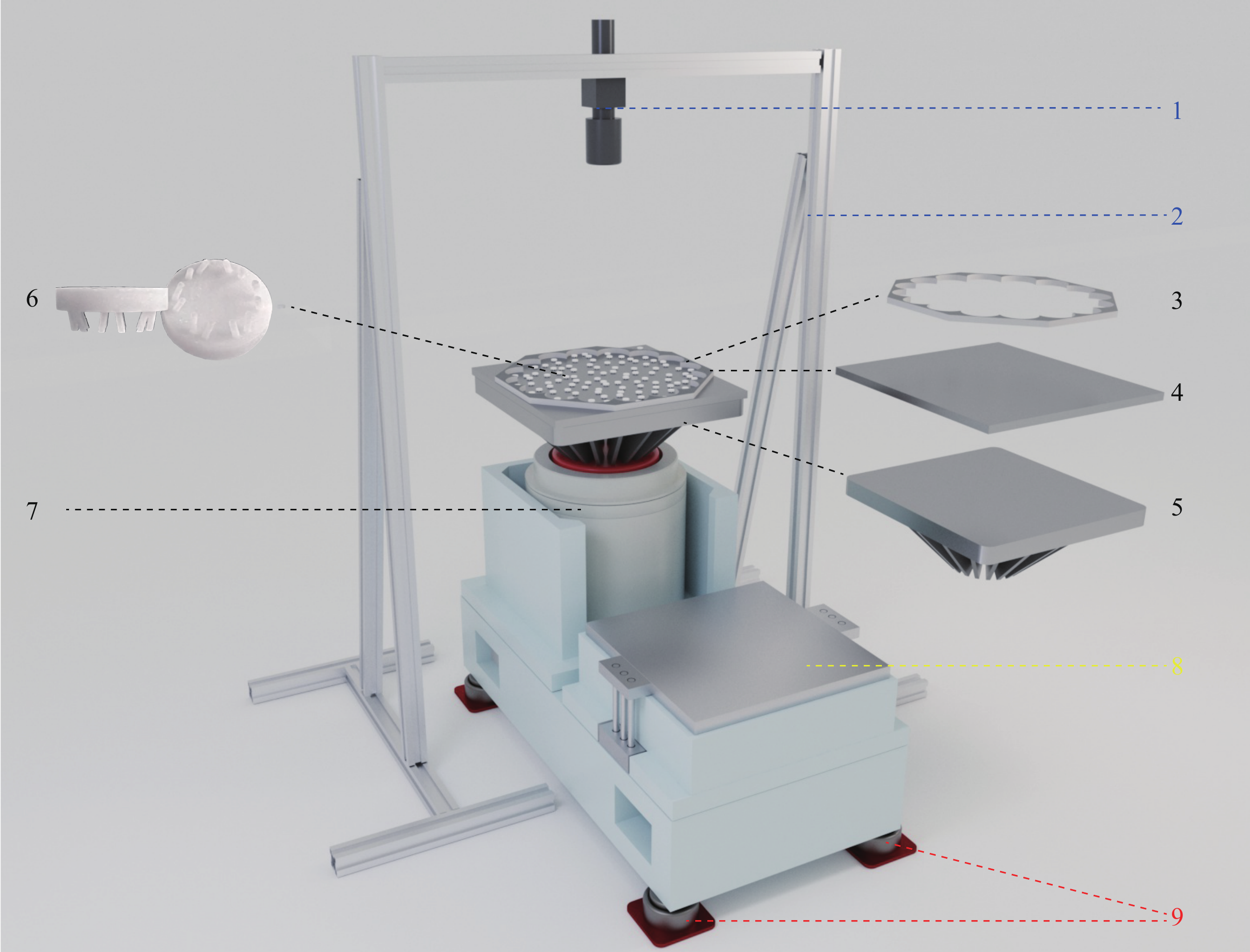}
\caption{\label{fig5} Schematic of experimental setup. 1.CCD camera;2.Aluminum profile frame;3.Flower-shape acrylic boundaries;4.Aluminum alloy plate; 5.Aluminum-magnesium alloy pedestal; 6. Brownian vibrators; 7.Electromagnetic vibrator; 8.Horizontal sliding stage; 9.Vibration isolation airbag}
\end{figure}

The schematic of the quasi-2D vibration experimental setup is shown in Figure~\ref{fig5}. The vibration driving system, indicated in black, consists of a vertically driven electromagnetic vibrator, an aluminum-magnesium alloy pedestal, an aluminum alloy plate, a flower-shape acrylic boundaries, and a horizontal layer of quasi-2d monodisperse Brownian vibrators. The horizontal adjustment system, marked in red, comprises four vibration isolation airbags located at the bottom of the setup. The optical imaging system, represented in blue, includes a CCD camera and related aluminum profile frame.

The electromagnetic vibrator, fixed to the aluminum-magnesium alloy pedestal, constitutes a vertical vibration platform that provides sinusoidal vibrations along the vertical direction. The aluminum-magnesium alloy pedestal weighs $60\si{kg}$ and has a truncated pyramid shape, with its upper surface being a $60\si{\centi\metre} \times 60\si{\centi\metre}$ plane. The unique microstructure of this aluminum-magnesium alloy provides appropriate vibration damping, reduces the amplification factor of high-frequency resonances, effectively suppresses potential surface standing waves, and thereby enhances the spatial uniformity of the vibration drive.
A flat aluminum alloy plate ($60\si{\centi\metre} \times 60\si{\centi\metre} \times 1\si{\centi\metre}$) is mounted onto the pedestal by eight strong F-clamps.
To prevent particle motion along the boundary, we utilize a flower-shaped acrylic confining boundary\cite{chen2022high}. Boundary particles consisting of three layers are excluded from the analysis. 

The Brownian vibrator, with a diameter of 16mm, is 3d-printed using CBY-01 resin with the density $\rho=1.14 \si{g/(\milli\metre)^3}$, Poisson's ratio $\nu=0.39$ and Young's modulus $Y=2  \si{GPa}$. Each Brownian vibrator features 12 legs. The legs are 3mm high and bent inward at an angle of $18.4^\circ$, alternately deviating from the mid-axis plane by $\pm 38.5^\circ$, as depicted in Figure~\ref{fig5}(d). Particle collisions are inelastic, with a restitution coefficient of approximately $\epsilon \approx 0.39$. The static friction coefficient between the particles and the base is approximately 0.42.
Due to the significantly longer time intervals between interparticle collisions compared to the persistent time (detailed in Fig ~\ref{sfig11} of Appendix F), the stochastic driving of the bottom surface on a given particle is minimally affected by the presence of neighboring particles.

The isolation airbags not only reduce the impact of mechanical vibrations on the optical imaging system but also serve as a means to adjust the levelness of the experimental platform. All data analyzed in this study were collected from experiments with continuous vibrations lasting over 20 hours, during which particles exhibited no significant gravity-driven drift. This ensures the isotropy of the driving force experienced by the particles. Precise adjustment and careful examination guarantee the stability of the experimental platform under vibration conditions, providing a solid foundation for data analysis.

To obtain the initial state at a specific packing fraction $\phi$, we randomly place particles and allow the system to run for two hours. The vibration is applied at a frequency of $f = 100$ Hz with a maximum acceleration of $a = 29.4$ m/s\textsuperscript{2}. The resulting amplitude, given by $A \equiv a/(2\pi f)^2 = 0.074$ mm, ensures a negligible vertical displacement of the particles. We capture particle configurations using a CCD camera at a frame rate of 40 frames/s for an hour, enabling further analysis and processing of the data.\par

\section{Discussion}\label{sec12}

It is curious why the granular and collective fluids were not observed previously\cite{olafsen1998clustering, Shattuck-PhysRevLett.98.188301, Olafsen-Urbach-2005, Pacheco-Vazquez-2009, Aranson-PRL-2000, Howell-PRE-2001, Oyarte-PRE-2013, Rivas-2011a, Rivas-2011b, Rivas-2012, Roeller-PRL-2011, Nahmad-PRE, Neel-PRE-2004, Clewett-2012, Luu-2013}, where a single or a shallow layer of spheres is confined within a quasi-2d cell and subject to vertical mechanical vibration.  In Ref.~\cite{Aranson-PRL-2000, Howell-PRE-2001, Oyarte-PRE-2013}, electrostatic and magnetic dipole forces introduce long-range interactions that differ substantially from our experiments. The experiments \cite{ Rivas-2011a, Rivas-2011b, Rivas-2012} focus on binary mixtures, different from ours. In Ref.~\cite{Shattuck-PhysRevLett.98.188301, Olafsen-Urbach-2005, Pacheco-Vazquez-2009, Roeller-PRL-2011, Nahmad-PRE, Neel-PRE-2004, Clewett-2012, Luu-2013}, there are only short-range repulsion and inelastic collisions between grains. There are some subtleties between our system and the previous experiments. The main issue is the randomization of particle motion at the single-particle level:  using a flat bottom plate \cite{ olafsen1998clustering, Roeller-PRL-2011, Nahmad-PRE, Neel-PRE-2004, Clewett-2012, Luu-2013} or a cover \cite{guan2021dynamics} introduces a non-Gaussian velocity distribution of a single particle, implying spatial correlations of particle movement, which cannot be eliminated with a rough plate or lid\cite{Olafsen-Urbach-2005, Shattuck-PhysRevLett.98.188301}.  The lack of Gaussian statistics in the single particle could induce phase separations\cite{olafsen1998clustering, Roeller-PRL-2011, Nahmad-PRE, Neel-PRE-2004, Clewett-2012, Luu-2013} due to a velocity-dependent energy injection rate \cite{Lobkovsky-Urbach-2009, Cafiero-Luding-2000}. The phase separation is absent when subject to random forcing in the simulation\cite{ Lobkovsky-Urbach-2009}, consistent with our observations. Before our experiments, two attempts were made to ensure Gaussian statistics of velocities\cite{Olafsen-Nature, scholz2017velocity}. However, additional effects were introduced, such as the motions of dimers and the continuous particle rotations along a single direction. Therefore, the Gaussian velocity in the single particle driving is crucial. 

In the 3d vertically vibrated hard sphere experiment \cite{scalliet2015cages} and accompanying simulations\cite{plati2019dynamical}, long-term superdiffusion of the detector ratchet angle was observed at high concentrations, which was thought to be linked to the asymmetric defects of the structure as suggested by a simplified simulation model to be associated with the direction of collective motion\cite{plati2022collective}. However, in our experiments, spontaneous collective behavior emerged due to the competition between active forces due to stochastic driving and the equivalent attraction due to inelastic collisions. Nevertheless, the direction of collective motion may be related to the inherent asymmetry in the experimental design, although it is not the root cause of the collective motion, as particles would exhibit collective motion at all $\phi's$ otherwise.

\section{Conclusion}\label{sec13}

Using Brownian vibrators, our study reveals the existence of four distinct phases in granular materials: granular fluid, collective fluid, poly-crystal, and crystal. Notably, the granular fluid phase showcases the intriguing behavior of purely-repulsive hard disks, where weak cohesion arises from inelastic collisions, resulting in liquid-like structures with fascinating dynamics. A remarkable large-scale collective motion emerges at $\phi=0.317$, coinciding with the peak in the fluctuations of local packing fractions. This collective motion persists until it terminates near $\phi=0.713$ where the system crystallizes. Our investigation demonstrates that granular materials subjected to uniform stochastic driving exhibit weak cohesion, complex internal structures, and, notably, a purely repulsive hard-disk system is capable of producing large-scale collective motion.

\backmatter

%\bmhead{Supplementary information}

%If your article has accompanying supplementary file/s please state so here. 

%Please refer to Journal-level guidance for any specific requirements.

\bmhead{Acknowledgments}

Y.C. and J.Z. acknowledge the NSFC (No. 11974238 and No. 12274291) support and the Innovation Program of Shanghai Municipal Education Commission under No. 2021-01-07-00-02-E00138. Y.C. and J.Z. also acknowledge the support from the Shanghai Jiao Tong University Student Innovation Center.

\noindent

\begin{appendices}

%In the Supplemental Materials, we provide more detailed descriptions of the following:\\
%(I) Voronoi diagrams at various values of $\phi$.\\
%(II) Two-dimensional (2d) maps showing the absolute value of the hexatic order parameter, $\left | \psi_6^j \right |$, at different $\phi$ values.\\
%(III) Global mean hexatic order and 2d correlations of the hexatic order parameter, $\psi_6^j$.\\
%(IV) Average curl of the particle displacement field.\\
%(V) Probability distribution functions (PDFs) of the particle displacement in the granular fluid, collective fluid, poly-crystal, and crystal, at different $\phi$ values.\\
%(VI) Radial distribution function, $g(r)$, and a comparison between the $g(r)$ of our experimental results, along with corresponding results for equilibrium hard disks.\\
%(VII) The P\'{e}clet number, $Pe$, and critical value, $Pe_c$.\\
%(VIII) Important timescales observed in the experiment.

\section{\label{sec:level1} The dynamics of active particles} 

From the theoretical work by Caprini et al.\cite{caprini2023flocking}, the dynamics of active particles reads:
\begin{align}
m\Dot{\mathbf{v_i}} =-\gamma \mathbf{v_i}+\mathbf{F_i}+\gamma_0 v_0\mathbf{n_i}+\sqrt{2\Lambda \gamma}\mathbf{\eta_i}, 
\end{align}
\begin{align}
\Dot{\mathbf{\theta_i}} =\sqrt{2D_r}\mathbf{\Theta_i}  
\end{align}

Our analysis focuses on the comparison of the dimensionless coefficients between the interparticle force $\mathbf{F_i}$ and the active force $\gamma_0 v_0\mathbf{n_i}$. First, by calculating the distribution of particle number density, we can determine the equivalent interaction force $\mathbf{F_i}$ of inelastic collisions between particles.

The radial distribution function, also known as the pair correlation function, is defined as: 
\begin{equation}\label{eq5}
g(r)= \frac{L^2}{2 \pi r \Delta r N(N-1)} \sum\limits_{j\neq k} \delta(r-|\vec{r}_{jk}|) ,
\end{equation}

where $L$ is the system size, $N$ is the total number of particles, $|\vec{r}_{jk}|$ refers to the relative distance between two particles $j$ and $k$, and $\Delta r$ is a small increment along the radial direction.
\par

\begin{figure*}[htpb]
\centering
\includegraphics[width=0.9\textwidth]{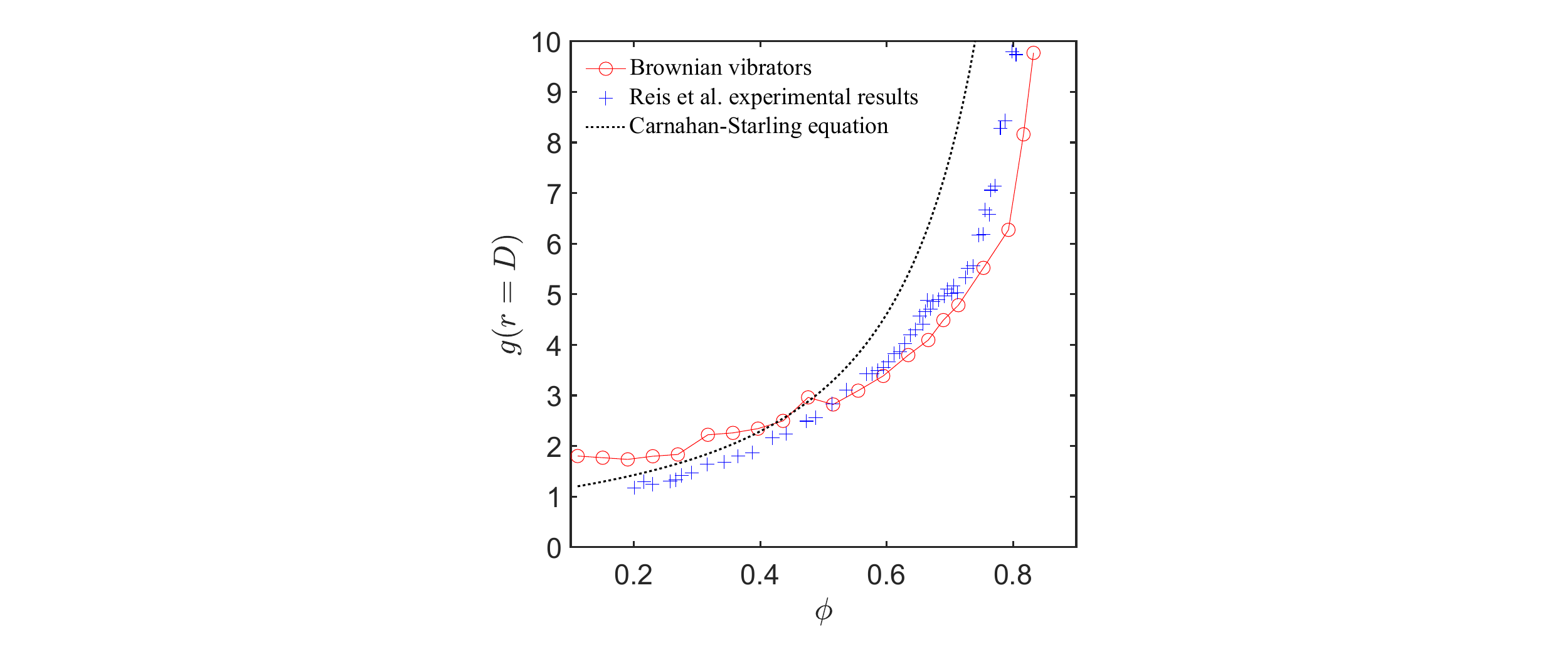}
\caption{\label{sfig1} $g(r = D)$ versus $\phi$. Red circles refer to our experimental results of Brownian vibrators. Blue pluses are the experimental results with vibrated stainless steel spheres in Ref.\cite{shattuck-PRL-crystal}. The black dotted line is the Carnahan-Starling equation. 
}
\end{figure*}

The particle configurations of our system are different from those of equilibrium hard disks and the previous experiment using vibrated stainless steel spheres Ref.\cite{shattuck-PRL-crystal}, as can be seen from the curve of 
$g(r=D)(\phi)\equiv g_{D}(\phi)$ shown in Fig.~\ref{sfig1} (a) with $D$ being the particle diameter. The black dotted line refers to the theoretical curve of Carnahan-Starling equation $g^{CS}_{D}(\phi)=[1-7/16\phi]/(1-\phi)^2$, which is derived from the equation of state for non-attracting hard spheres\cite{carnahan1969equation}. \par

Assuming that the inelastic collisions between particles can be effectively represented by an interparticle attractive potential, the radial distribution function is given by $g(r) = \exp(-V(r)/(kT)) \cdot g_{\text{eq}}(r)$, where $V(r)$ represents the effective potential and $kT\equiv\frac{1}{2}m v_T^2$ can be estimated from the thermal velocity of the particles, denoted as $v_T$. The definition of the thermal velocity is provided below. The equilibrium radial distribution function, $g_{\text{eq}}(r)$, for purely repulsive hard spheres is obtained from the BGY theory as described in Ref.~\cite{chae1969radial}. The interparticle force is given by $F_i= -\frac{d(V(r))}{dr}$, and its maximum value is determined by:

\begin{align}
F_{\text{max}} = \frac{1}{2}m v_T^2 \left[\frac{d\ln\left(\frac{g(r)}{g_{\text{eq}}(r)}\right)}{dr}\right]_{\text{max}}.
\end{align}

The thermal velocity, $v_T$, can be obtained by calculating the ensemble-averaged root mean square (RMS) velocity across different packing fractions, $\phi$, using the formula:
\begin{align}
v_T=\sqrt{\frac{\sum_{i=1}^{N_{\phi}} v_i^2}{N_{\phi}}},
\end{align}

where $N_{\phi}$ represents the total number of particles at a given packing fraction $\phi$, and $v_i$ denotes the RMS velocity of a specific particle $i$. Here the thermal velocity $v_T=1.4\pm 0.1D/s$ is weakly affected by $\phi$ \cite{chen2022high}.

Thus, we have completed the calculation of the equivalent attractive force of inelastic collisions at different packing fractions. Next, we will introduce the calculation details of the P\'{e}clet number $Pe$ at different packing fractions.

\begin{figure}[htpb]
\centering
\includegraphics[width=0.45\textwidth]{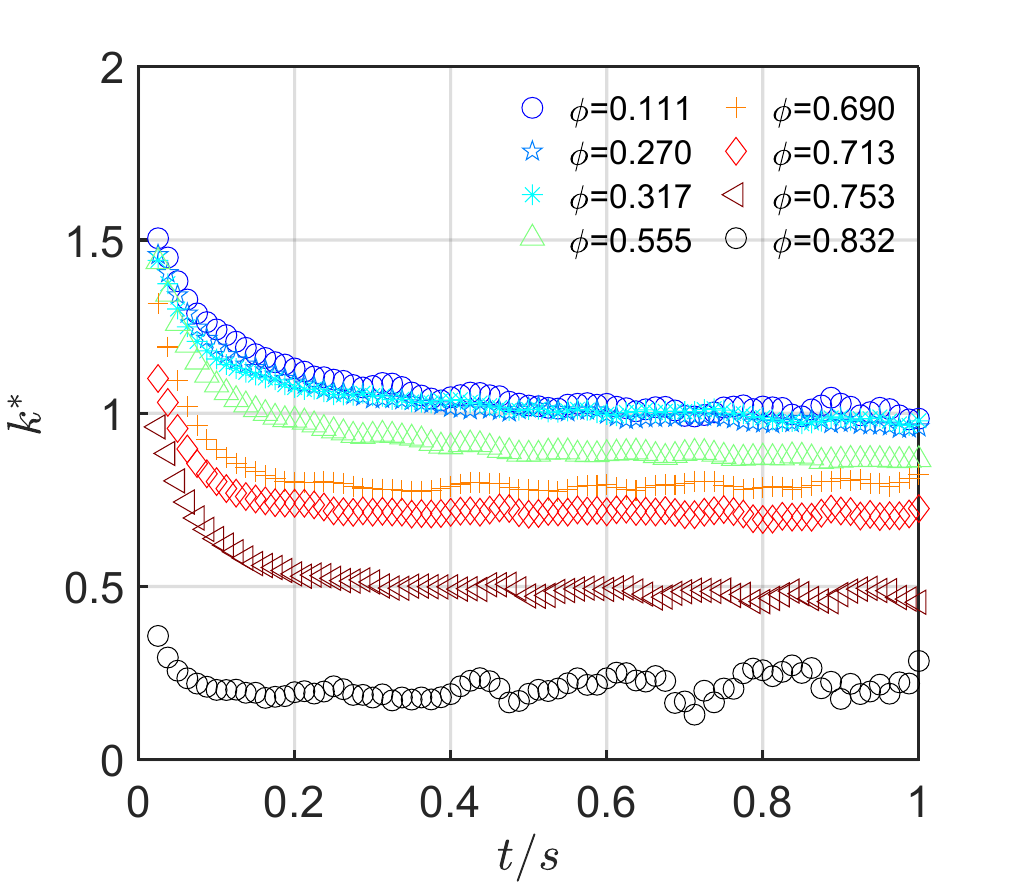}
\caption{\label{sfig2} The local slope of the mean square displacement (MSD) curve in double logarithmic coordinate $k^*=\frac{d(log(MSD))}{d(log(t))}$.
}
\end{figure}

\begin{figure}[htpb]
\centering
\includegraphics[width=0.45\textwidth]{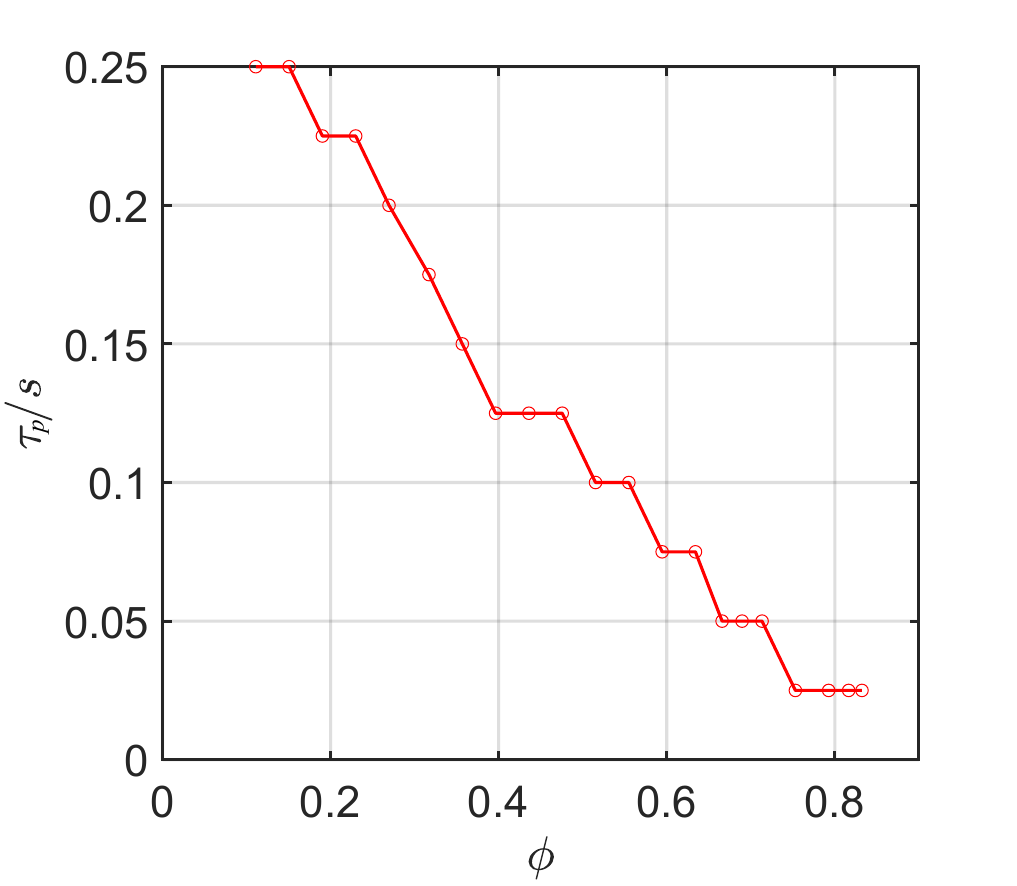}
\caption{\label{sfig3} The persistence time $\tau_p$. The stair-like structure is due to the finite time resolution of 0.025 s. Since the experimental frame rate is 40 frames per second, the time resolution is 0.025s.
}
\end{figure}

In our experiment, the persistence time, denoted as $\tau_p$, can be determined from the cutoff time corresponding to the initial segment of superdiffusion. This segment is identified by the condition where the local slope of the mean square displacement (MSD) curve, represented by $k^*=\frac{d(\log(MSD))}{d(\log(t))}$ in double logarithmic coordinates, is greater than 1, as illustrated in Fig.~\ref{sfig2}.
Considering the fluctuations of the measured slope $k^*$ in the experiment, where $k^*= 1 \pm 0.1$ within the time scale of the random walk, we can determine $\tau_p$ as $\tau_p=\max(t(k^*>1.1))$. This maximum value of $t$ is depicted in Fig.~\ref{sfig3}.

To determine the root mean square of the active velocity, denoted as $v_0$, single-particle experiments are conducted. The estimation of $v_0$ is obtained as follows:

\begin{align}
v_0=\sqrt{\langle \overline{ \mathbf{v}^2(t)}\rangle} \approx 1.52  \si{D/s} \approx 24.3  \si{mm/s}
\end{align}

Here, the notation $\langle \cdot \cdot \cdot\rangle$ represents the average over repeated experiments, and $\overline{ \cdot \cdot \cdot }$ indicates the time average over the particle trajectory. 

The persistence length of particles, denoted as $v_0\tau_p$, quantifies the distance over which a particle maintains directional persistence. It represents the range within which its motion remains coherent before being randomized by random forces. And the P\'{e}clet number $Pe \equiv v_0/(D_rD)$ is viewed as the ratio between persistence length and particle size which in this case is D=16 mm. Here the persistence time $\tau_p=1/D_r$ \cite{caprini2023flocking}. Therefore, theoretically, $Pe$ can also be calculated through the definition. However, due to the higher spatial resolution precision in the experiment (0.29mm per pixel) compared to the temporal resolution precision (average particle displacement of 0.6mm within 0.25 seconds), the $Pe$ presented in the main text is obtained from the corresponding mean square displacements over the persistent time.

Consider the motion of an individual Brownian particle on an aluminum alloy plate. When the vibrational driving is stopped, the time required for the particle with an average velocity $v_0$ to decelerate to 0 due to friction is $\tau_f=v_0/(\mu g)$, where $\mu$ is the friction coefficient and $g$ is the acceleration due to gravity. $\mu$ has been measured to be approximately 0.420. This measurement was obtained by determining the angle of repose through an experimental setup involving a gradual and continuous tilting of a slope from its horizontal position. The angle of repose is identified as the point where the particle exhibits incipient instability and starts to slide. Then we obtain:
\begin{align}
\tau_f=\frac{v_0}{\mu g}\approx 5.91 \si{ms}. 
\end{align}

In our vibration experiments, the motion of particles can be classified into two types: (a) collisions with the vibrating platform, and (b) a projectile-like leap motion where particles momentarily detach from the platform. The inertia time, denoted as $\tau_I$, of an individual vibrating particle is determined by two factors: the time required for the velocity to be completely dissipated through multiple collisions with the platform, and the cumulative duration of all the leap periods between these collisions. Mathematically, it can be expressed as:

\begin{align}
\tau_I = \frac{\tau_f}{\tau_c}(\tau_c+\tau_h).
\end{align}

Here, $m$ represents the particle mass, $\gamma$ is the drag coefficient, $\tau_c$ is the contact time for mutual collisions of spheres with radius $R$, and $\tau_h=38ms$ denotes the average duration of a single particle's leap, which is measured using a high-speed camera. The contact time $\tau_c$ can be calculated using a formula derived from the reference \cite{landau1986theory, schwager2008coefficient, scholz2017velocity}:

\begin{align}
\tau_c= 3.218R\frac{[\sqrt{2}\pi\rho(1-\nu^2)]^{2/5}}{Y^{2/5}v_c^{1/5}}
\end{align}

In our experiment, a single collision between our Brownian vibrators and the platform can be effectively approximated as a collision between one of the 12 inclined legs and the platform. Considering that the legs of the particles are cylindrical with a radius of 0.5mm, we can treat the collision as a mutual collision between spheres with a radius R=0.5mm for calculation purposes. 

Our Brownian vibrators are 3d-printed using CBY-01 resin.
JS-UV-CBY-01 Light Yellow High-Temperature Hard Material is an ABS-Like Stereoscopic Modeling Resin with precise and durable properties, which is used in solid-state laser light curing molding.
At a temperature of 25 $\si{{\degreeCelsius}}$, the density$\rho=1.14 \si{g/(\milli\metre)^3}$, Poisson's ratio: $\nu=0.39$, Young's modulus: $Y=2  \si{GPa}$, the viscosity is 400 $\si{\milli Pa \cdot s}$. In our 3d-printing process, we utilized a curing depth of 0.135 mm and a construction layer thickness of 0.05 mm.
The collision velocity, denoted as $v_c$, can be estimated as $v_c=\frac{\tau_h g}{2}\approx 186mm/s$, where $\tau_h$ is the average duration of a single particle's leap and $g$ represents the acceleration due to gravity. The maximum velocity of the vibrating platform along the z-axis is 46.8mm/s. Since the collision velocity term $v_c$ in the equation above is raised to the power of $1/5$, the correction for the platform motion's impact on the collision velocity can be neglected in the calculation.

By substituting the above parameters into equation (A9), we obtain $\tau_c \approx 0.193 \si{ms}$. Then, by plugging $\tau_c$ into equation (A8), we get $\tau_I \approx 1.17 \si{s}$.

The magnitude of the equivalent dissipative force is given by $\left | \mathbf{F_d} \right | \equiv mv_0/\tau_I$. After reaching a steady state, the energy dissipated by the particle is equal to the energy input from the active force. Therefore, the magnitude of the active force $\left | \mathbf{F_a} \right |=\left | \mathbf{F_d} \right | \equiv mv_0/\tau_I= \gamma_0 v_0$, where $\gamma_0$ is the equivalent drag factor, given by $m/\tau_I$, and $m$ is the mass of the particle.

Here the magnitude of active force $\gamma_0 v_0$ is generated through the friction between the particle and the platform due to the unique structure of the inclined legs of the particles, which is different with the simulation\cite{caprini2023flocking}. However, this distinction does not significantly impact the relevant analysis presented in the main text, as it primarily involves comparing the dimensionless factors associated with the second and third terms.

By comparing the maximum equivalent attractive force and the active force, the critical value can be evaluated.

\begin{align}
    Pe_c=F_{max}/(\gamma v_0) \cdot Pe= \frac{1}{2}m v_T^2 \left[\frac{d\ln\left(\frac{g(r)}{g_{eq}(r)}\right)}{dr}\right]_{max} /(\gamma v_0) \cdot Pe \equiv \lambda Pe
\end{align}

The dimensionless coefficient, denoted as $\lambda$, exhibits an expected increase with an increase in $\phi$. This trend is consistent with the fact that the effective attraction force also increases with $\phi$ due to the increasing collision frequency. For instance, at $\phi=0.270$, $\lambda$ is approximately 0.72, while at $\phi=0.317$, $\lambda$ reaches around 1.04. These values indicate that when the maximum effective attractive force surpasses the active forces, collective behavior starts to emerge according the theoretical predictions\cite{caprini2023flocking}.
\newpage

\section{\label{sec:level1} Voronoi diagrams} 
\begin{figure*}[htpb]
\centering
\includegraphics[width=0.9\textwidth]{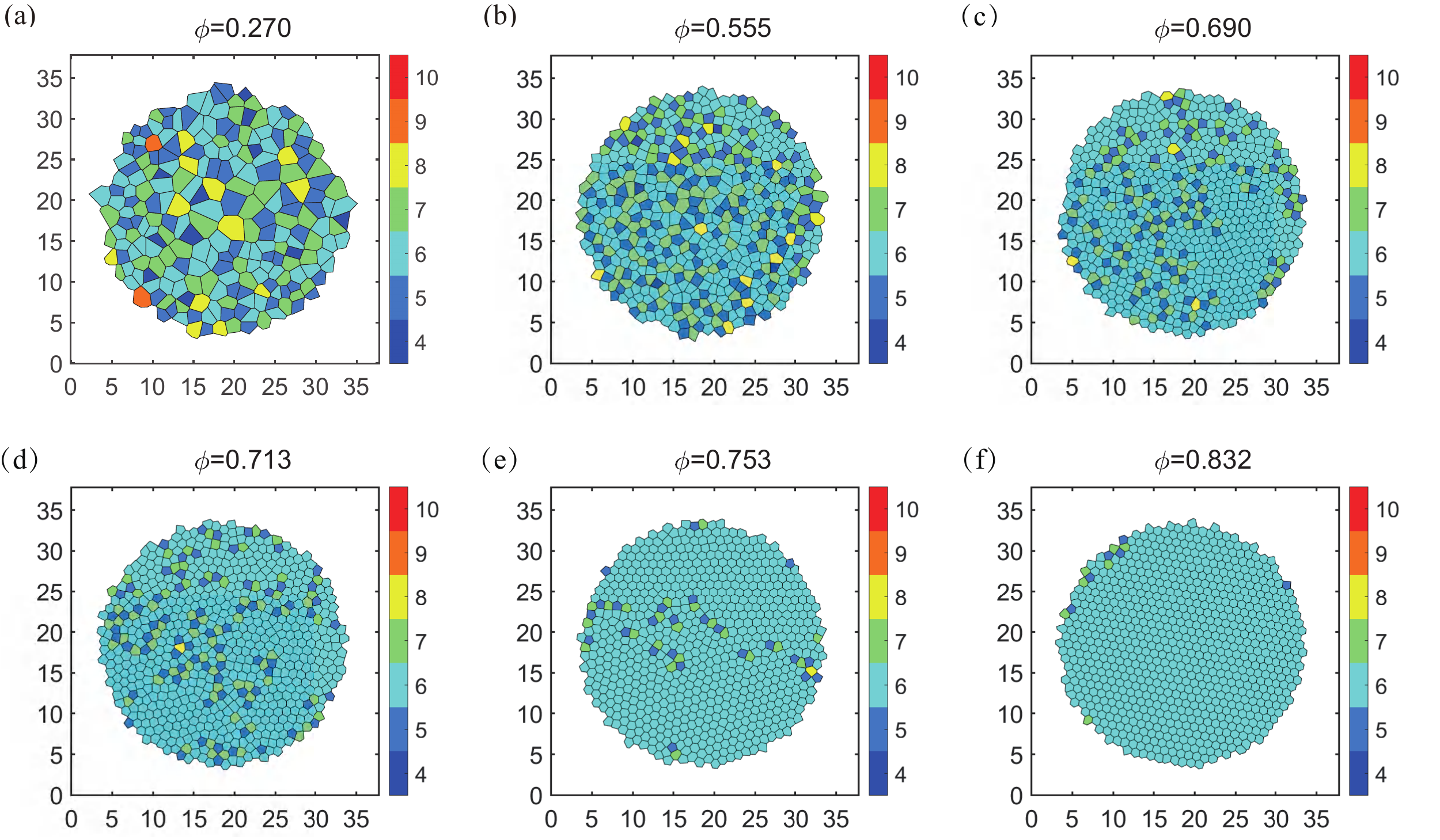}
\caption{\label{sfig4} (a-f) Voronoi diagrams at various $\phi's$.
}
\end{figure*}
Figure~\ref{sfig4} illustrates the Voronoi diagrams at different values of $\phi$. In the granular fluid phase ($\phi=0.270$), the Voronoi cell order, represented by the number of edges, ranges from 4 to 9 and exhibits a random spatial distribution. In the collective fluid phase ($\phi=0.555$), the order ranges from 4 to 8, and there is no noticeable localized ortho-hexagonal aggregation. In comparison, in the poly-crystalline phase ($\phi=0.713$), the majority of Voronoi cells are of order six, and the grain boundaries consist of pairs of grains with orders 5 and 7. This observation holds true for $\phi=0.753$ as well. With the exception of defects near boundaries, the crystalline region at $\phi=0.832$ comprises Voronoi cells of order six that span the entire system.
By utilizing the area $S_v$ of Voronoi polygons, we can calculate the local packing fraction $\phi_{local}=\frac{\pi D^2}{4 S_v}$  and analyze its distribution and fluctuation.

\section{\label{sec:level1} The hexatic order parameter} 

The hexatic order parameter, $\psi_6^j$, quantifies the similarity between the structure surrounding a particle $j$ and a perfect hexagonal arrangement. It is defined as $\psi_6^j=\frac{1}{n_j}\sum_{k=1}^{n_j}e^{i6\theta_{jk}}$, where $n_j$ represents the number of nearest neighbors of particle $j$, and $\theta_{jk}$ denotes the angle between the vector $\vec r_{jk}\equiv\vec r_k- \vec r_j$ and the x-axis. Notably, $n_j$ is not restricted to six; it can take different values, such as 2 for a linear structure, 3 for a trident structure, 4 for a center particle with four neighbors, and so on.

To illustrate, let's consider the case where $n_j=2$. When the center particle and its two neighbors form a straight line, $\psi_6^j$ equals 1. Consequently, $\psi_6^j$ characterizes the local particle-scale symmetry.

\begin{figure*}[htpb]
\centering
\includegraphics[width=0.9\textwidth]{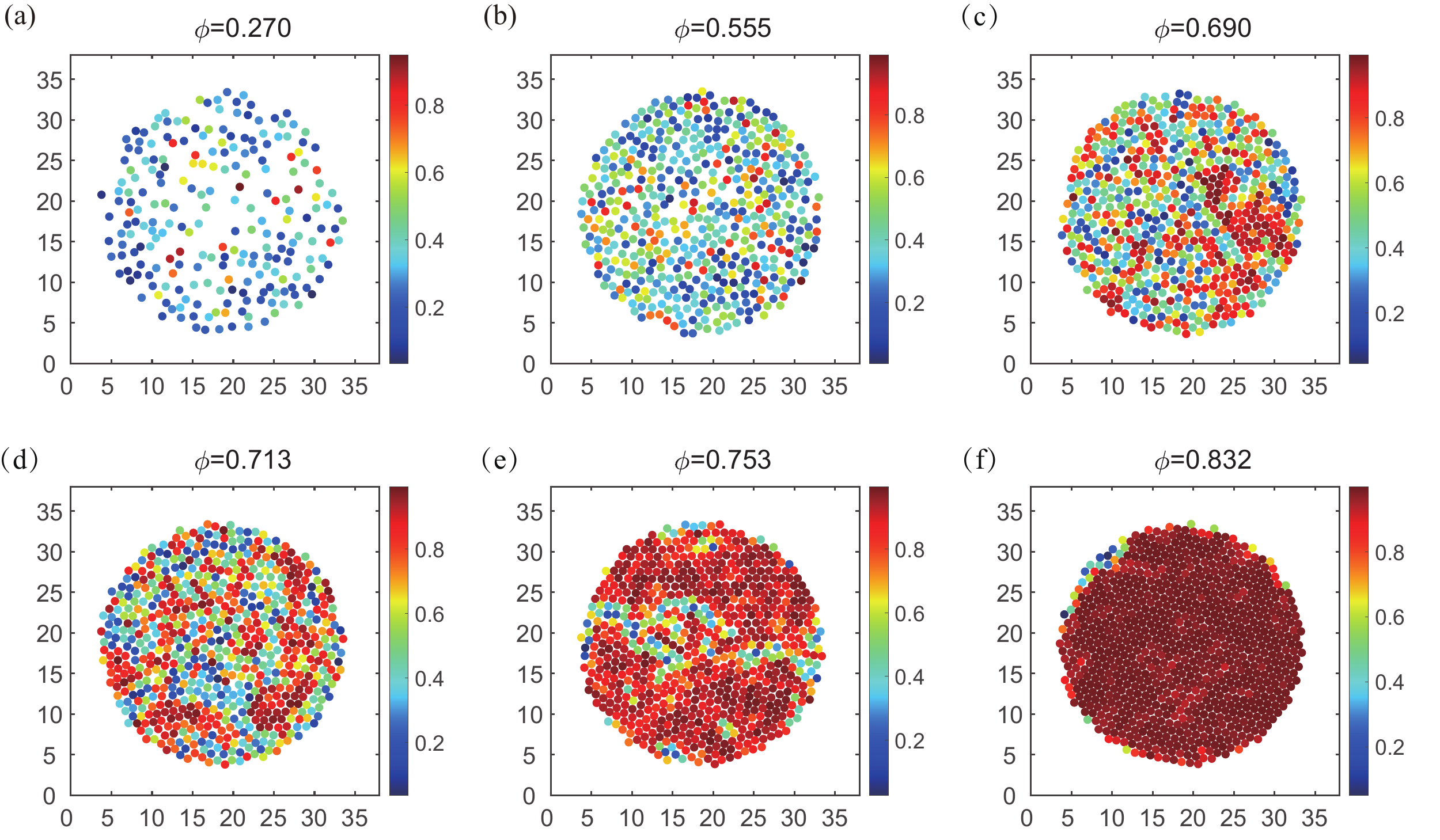}
\caption{\label{sfig5} (a-f) The absolute value of hexatic order parameter $\left | \psi_6^j \right |$ at various $\phi's$.
}
\end{figure*}

The 2d maps displaying the absolute value of the hexatic order parameter, $\left | \psi_6^j \right |$, at various $\phi$ values are presented in Fig.~\ref{sfig5}. Each panel in the figure corresponds to the same particle configurations as shown in Fig.~\ref{sfig4}.

\begin{figure}[htpb]
\centering
\includegraphics[width=0.45\textwidth]{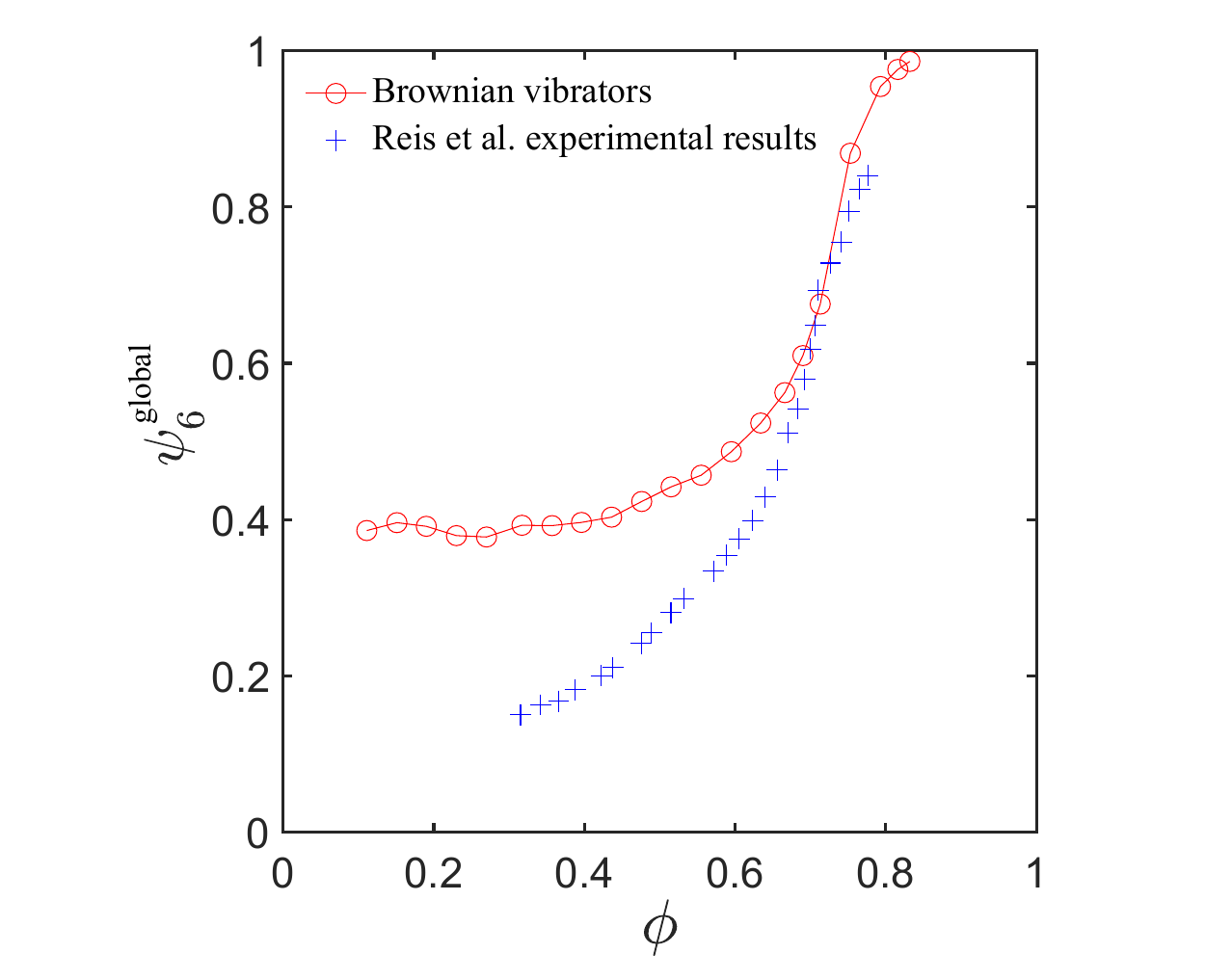}
\caption{\label{sfig6} The global mean of $\psi_6$ at various $\phi's$. Red circles refer to Brownian vibrators. Blue pluses are the results of Ref.\cite{shattuck-PRL-crystal}.
}
\end{figure}

\begin{figure*}[htpb]
\centering
\includegraphics[width=0.9\textwidth]{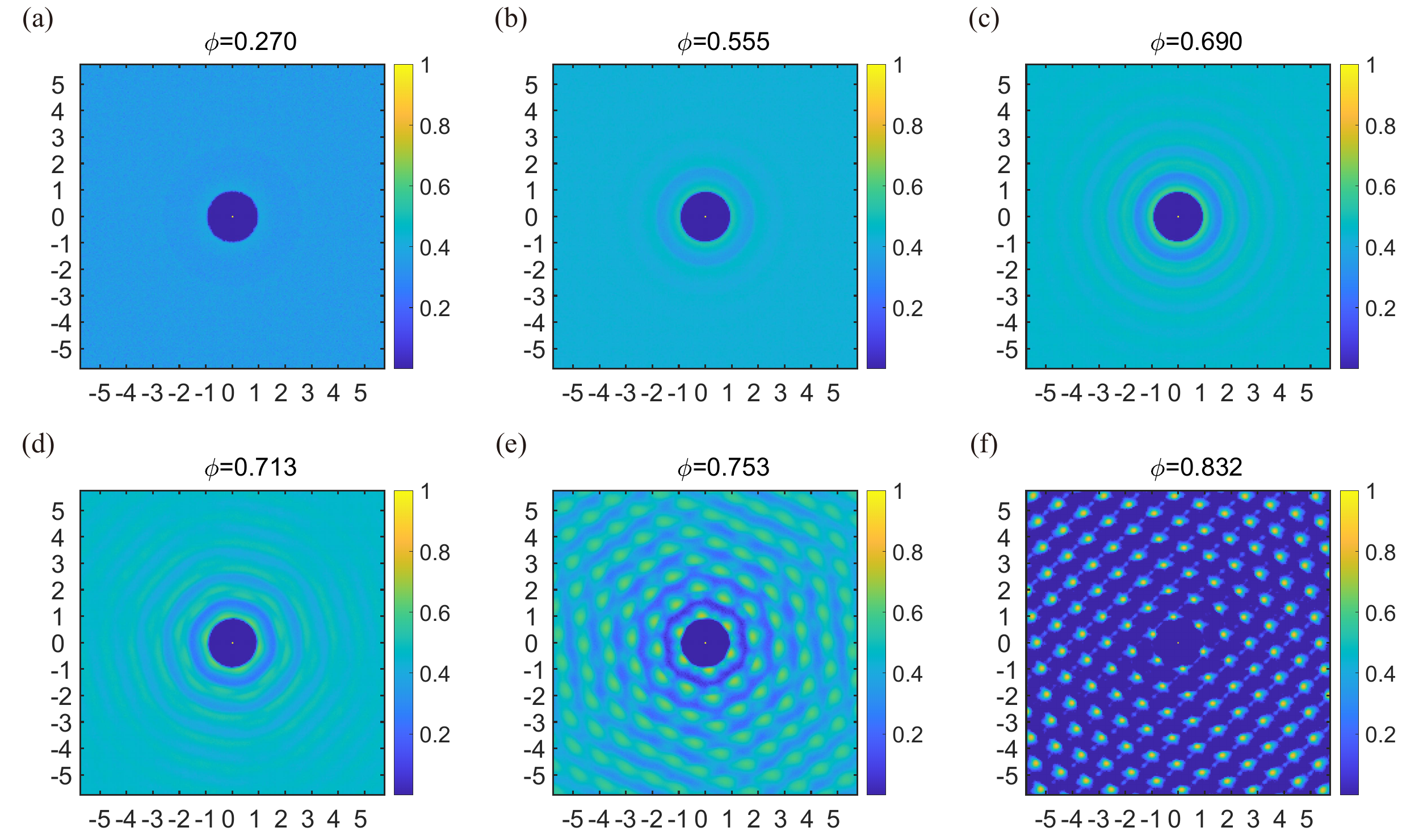}
\caption{\label{sfig7} (a-f) The 2d correlations of the local hexatic order $g^{2D}_6(r,\theta)$ at various $\phi's$.
}
\end{figure*}

The global mean hexatic order parameter, $\psi_6^{global}= \langle \vert \frac{1}{N} \sum_{j=1}^N \psi_6^j \vert \rangle$, is calculated by averaging the absolute value of the hexatic order parameter for all particles over different configurations. Here, $N$ represents the total number of particles, and $\langle ... \rangle$ denotes the average over different particle configurations.

In Fig.~\ref{sfig6}, we compare the global mean $\psi_6^{global}$ observed in our experiment with the results from Ref.\cite{shattuck-PRL-crystal}, particularly showing a plateau for packing fractions $\phi \le 0.436$.

Moving on to Fig.~\ref{sfig7}, we present the two-dimensional correlations of the local hexatic order, denoted as $g^{2D}_6(r,\theta)= \frac{L^2}{r\Delta r \Delta \theta N(N-1)} \sum\limits_{j\neq k} \delta(r-|\vec{r} _{jk}|) \delta(\theta-|\theta_{jk}|) \psi^j_6 \psi^k_6 $. It is evident from the figure that $g^{2D}_6(r,\theta)$ exhibits isotropic behavior for $\phi \le 0.690$, while it demonstrates hexagonal symmetry for $\phi\ge0.713$, indicating clear crystalline characteristics.

\section{\label{sec:level1} The PDFs of particle displacement} 

\begin{figure*}[htpb]
\centering
\includegraphics[width=0.9\textwidth]{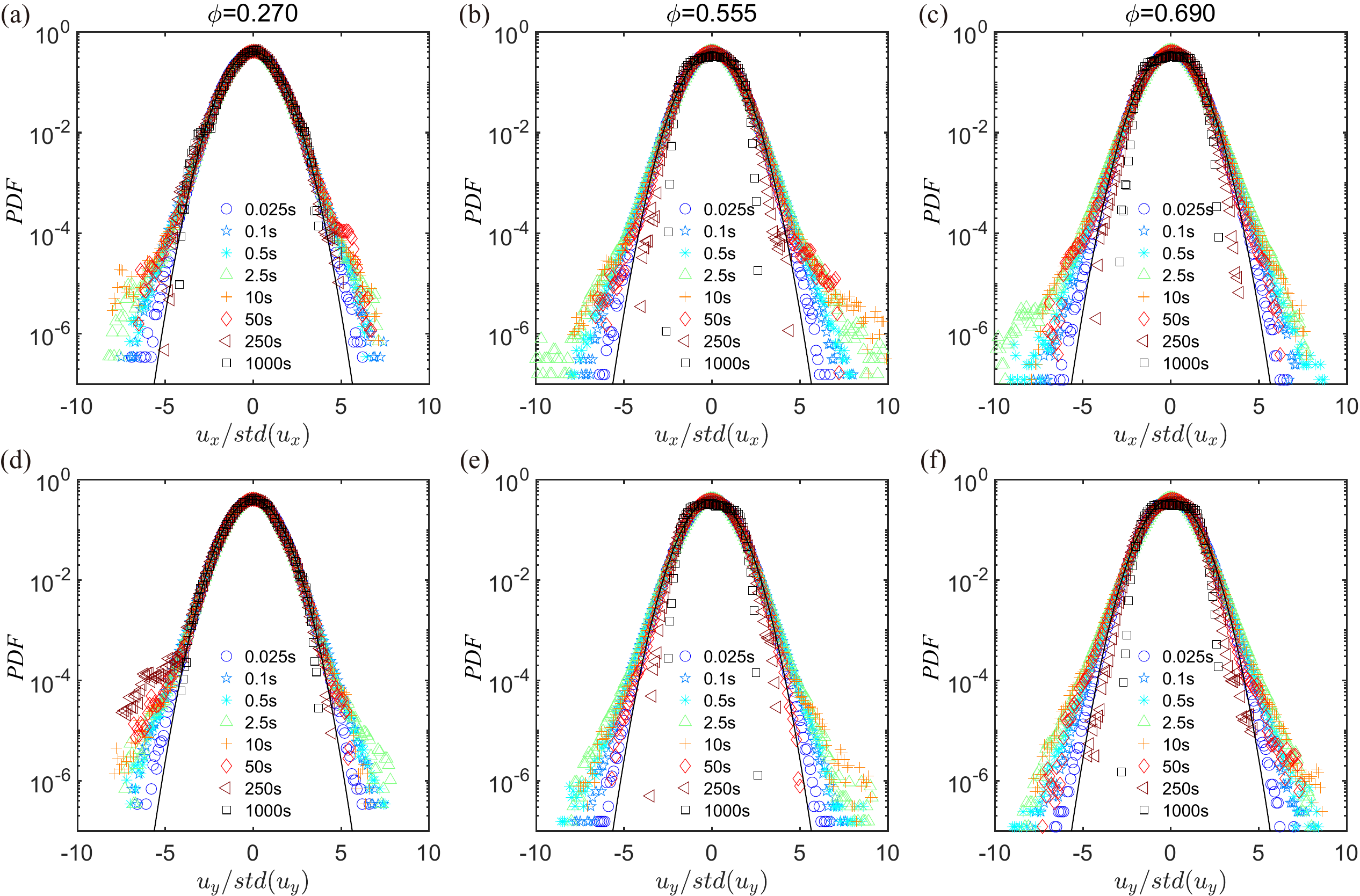}
\caption{\label{sfig8} The PDFs of $u_{\alpha}/std(u_{\alpha})$ ($\alpha$ refers to the $x$ or $y$ component) at the volume fractions of $\phi=0.270,0.555,0.690$. Different colors and markers represent various time intervals. The black solid lines refer to Gaussian distributions. (a-c) $u_x/std(u_x)$ (d-f)$u_y/std(u_y)$.}
\end{figure*}

\begin{figure*}[htpb]
\centering
\includegraphics[width=0.9\textwidth]{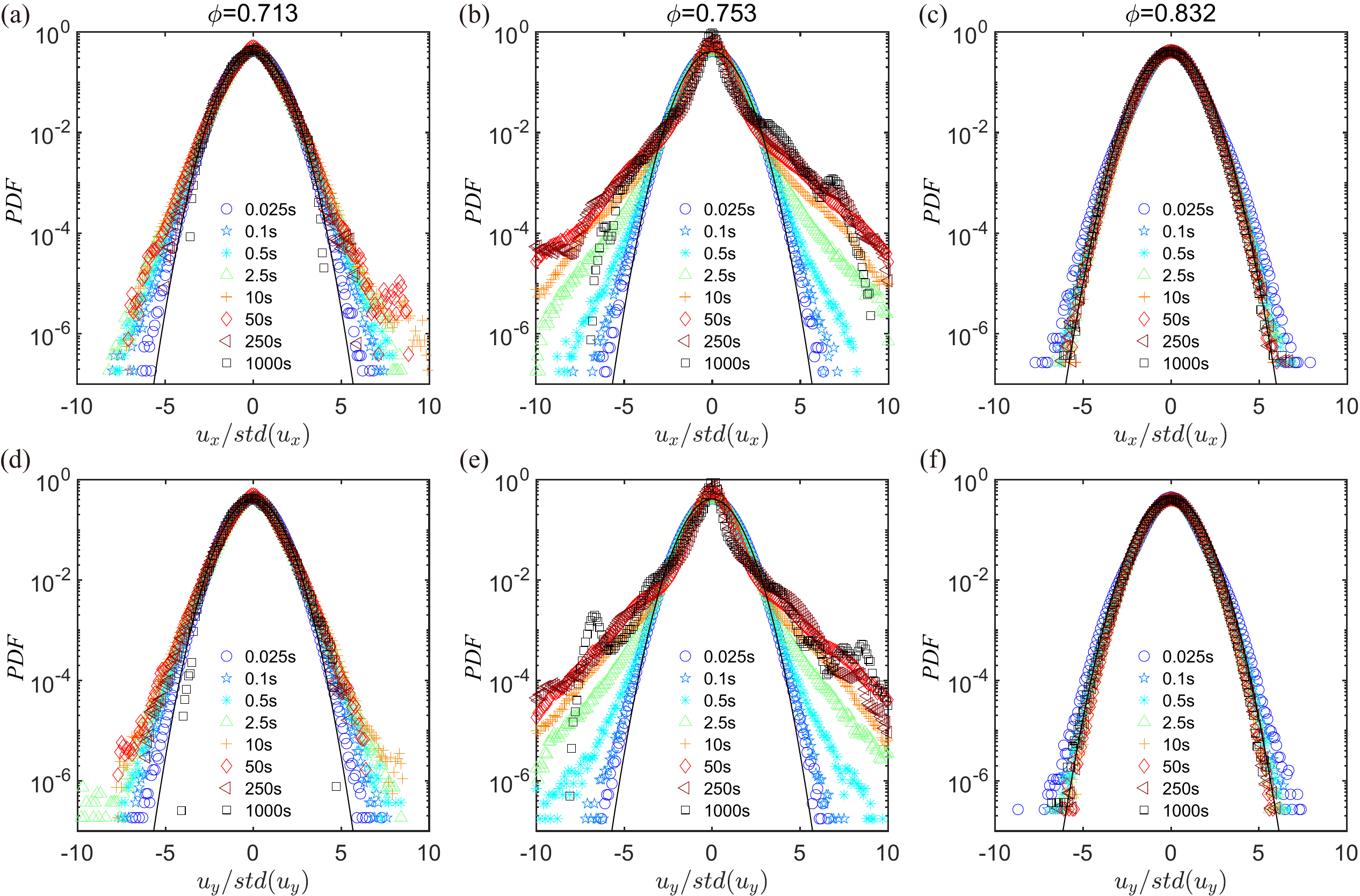}
\caption{\label{sfig9} The PDFs of $u_{\alpha}/std(u_{\alpha})$ ($\alpha$ refers to the $x$ or $y$ component) at the volume fractions of $\phi=0.713,0.753,0.832$. Different colors and markers represent various time intervals. The black solid lines refer to Gaussian distributions. (a-c) $u_x/std(u_x)$ (d-f)$u_y/std(u_y)$.}
\end{figure*}

The probability density functions (PDFs) of $u_x$ and $u_y$, representing the x and y components of the particle displacement vector $\vec u$, are presented in Fig.~\ref{sfig8} and Fig.~\ref{sfig9}, respectively. In these figures, the horizontal coordinates are normalized by the corresponding standard deviation of particle displacement. The solid black lines in the figures depict Gaussian distributions for comparison.

Notably, none of the PDF curves exhibit a Gaussian shape. As $u_x/std(u_x)$ or $u_y/std(u_y)$ varies above and below zero, each curve progressively deviates from the Gaussian distribution. However, in most cases, except for $\phi=0.753$, the Gaussian distributions can still offer an approximate description of the PDFs, particularly for the central part of the distributions. It is primarily in the tails of the PDFs, where the probabilities are less than $10^{-3}$, that the Gaussian approximation becomes less accurate.

In the case of the volume fraction $\phi=0.753$, the PDFs of $u_x/std(u_x)$ and $u_y/std(u_y)$ exhibit the most significant deviations from Gaussian behavior at long time intervals. This deviation is attributed to the coexistence of a large number of nearly perfect crystal particles and a relatively small number of grain boundary particles, which leads to distinct features in the PDF distributions.

\begin{figure*}[htpb]
\centering
\includegraphics[width=0.9\textwidth]{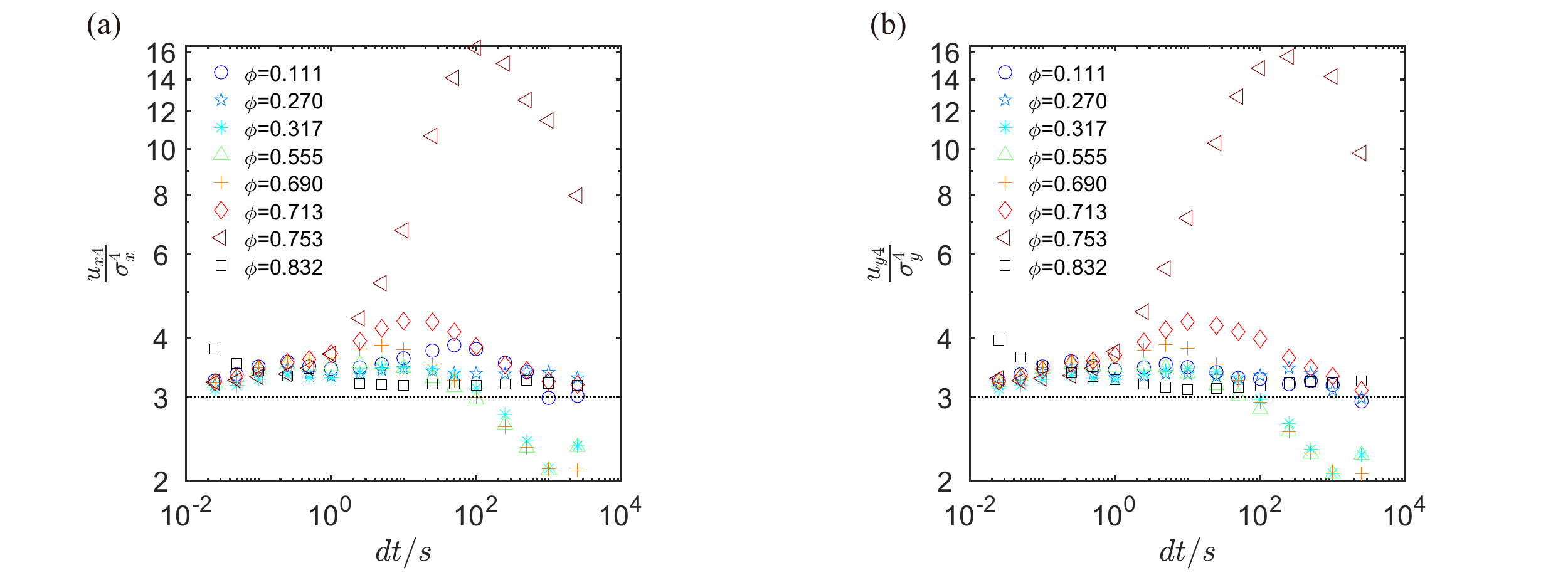}
\caption{\label{sfig10} The kurtosis of the PDFs of $u_x/std(u_x)$ (a) and $u_y/std(u_y)$ (b) for a series of time intervals and different packing fractions $\phi's$. }
\end{figure*}

The departure of the PDFs from Gaussian behavior can be quantitatively assessed using the kurtosis, as depicted in Fig.~\ref{sfig10}. The kurtosis is defined as follows:

\begin{equation}\label{eq4}
\frac{\mu_4}{\sigma^4}=\frac{\frac{1}{n}\sum_{i=1}^n(u_i-\bar{u})^4}{(\frac{1}{n}\sum_{i=1}^n(u_i-\bar{u})^2)^2},
\end{equation}

where $u_i$ represents a random variable, which can refer to either $u_x/std(u_x)$ or $u_y/std(u_y)$. It is important to note that the kurtosis of a Gaussian distribution equals 3.

\section{\label{sec:level1} The average curl of the particle displacement field}
We focus on the particle displacement field to quantify the large-scale collective motion.
First, the particle density field can be coarse-grained with the kernel function $\Psi(\vec r)=e^{-(2\vec r/D)^2}$:
\begin{equation}\label{eq1}
\rho(\vec r)=\sum_{i=1}^N \Psi(\vec r-\vec r_i). 
\end{equation} 
Next, the particle displacement field can be coarse-grained within a time interval $\Delta t$:
\begin{equation}\label{eq2}
\vec u(\vec r,\Delta t)=\rho(\vec r)^{-1} \sum_{i=1}^N \vec u_i(\Delta t) \Psi(\vec r).
\end{equation}
Here, the displacement of the particle $i$ is $\vec u_i(\Delta t)=(\vec r_i(t_0+\Delta t)-\vec r_i(t_0))$, and the collective behavior is evident in Fig. 1 (e) in the main text for $\Delta t=100s$.
In a quasi-two-dimensional system, the average curl of the displacement field $\Omega$ is defined as:
\begin{equation}\label{eq3}
\begin{split}
\Omega&=\langle \left[ \nabla \times \vec u(\vec r)\right]_z \rangle \\
&=\langle \rho(\vec r)^{-2} \sum_{i=1}^N\sum_{j=1}^N \Psi(\vec r)_j \left[ \nabla \Psi(\vec r)_i \times \vec u_{ij} \right]_z \rangle.
\end{split}
\end{equation}
Here, the angle brackets represent the spatial average followed by a time average over different $t_0$.

\section{\label{sec:level1} Important timescales in the experiment} 
 \begin{figure*}[htpb]
\centering
\includegraphics[width=0.45\textwidth]{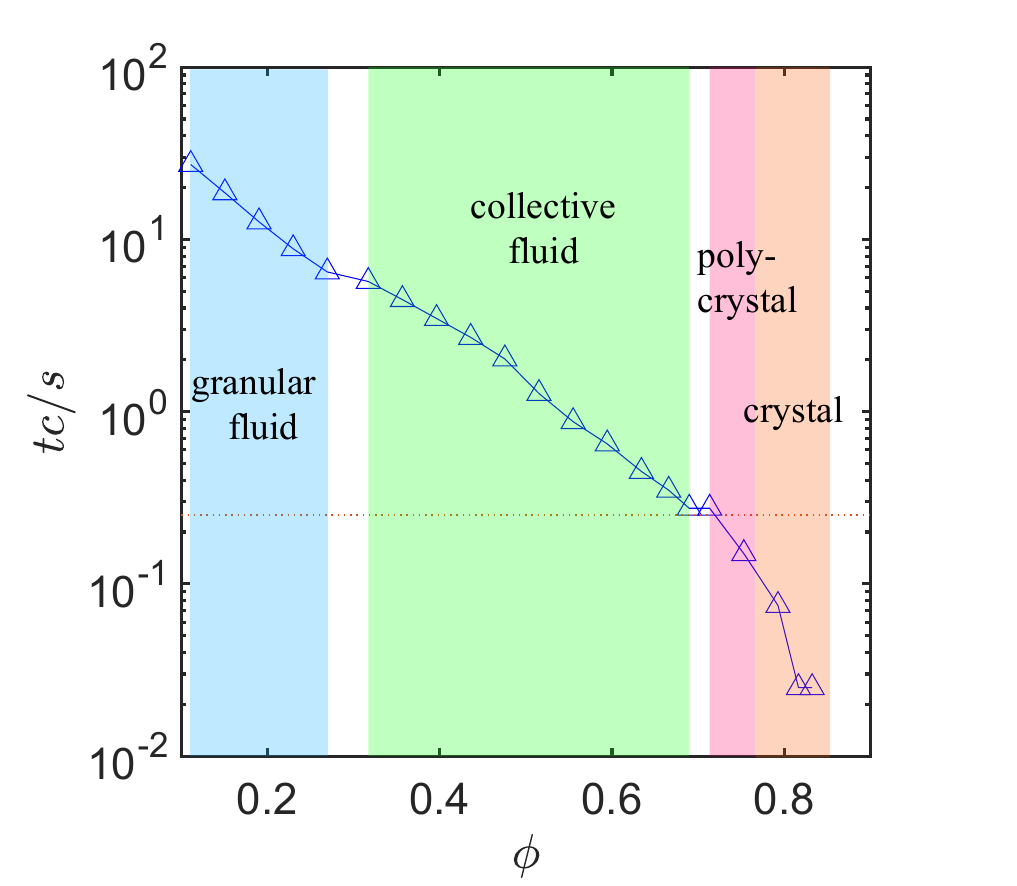}
\caption{\label{sfig11} The mean free time of particles was measured at various packing fractions. A red dotted line marks the characteristic time t=0.25s, representing the onset of the single-particle's diffusive motion.
}
\end{figure*}

Our experiment involves four crucial timescales:

(1) The time interval of stochastic driving, estimated by the vibration frequency of the electromagnetic shaker, is approximately $1/f=0.01$ seconds.

(2) The persistence time, denoted as $\tau_p$, is obtained from the particle's mean square displacement (MSD) curve and is around 0.25 seconds for a single particle system.

(3) The particle's mean free time, denoted as $t_c$, is illustrated in Figure~\ref{sfig11}. Here, $t_c(\phi)$ is defined as the time when the MSD of particles at a given packing fraction $\phi$ reaches the mean separation distance of neighboring particles, which is equal to $(\sqrt{\pi/(2 \sqrt{3}\phi)}-1) D$. In the granular fluid phase ($0.111 \leq \phi \leq 0.270$) and the collective fluid phase ($0.317 \leq \phi \leq 0.690$), which are the most interesting regimes in terms of physics, the mean free time between particle-particle collisions is much larger than 0.25 seconds. This indicates that stochastic driving randomizes the particle's velocities well before two adjacent particles interact.

(4) The period of collective motion around the system's center is approximately 4000 seconds. In the collective fluid phases, the MSD peaks around 2000 seconds correspond to half a rotation period of the outermost particles' collective motion. These particles contribute the most to the MSD, as their displacement follows an approximately semicircular path from one side of the system to the other, covering a distance of about 30 particle diameters. Once the displacement reaches this magnitude, the MSD starts to decrease.

\end{appendices}


\bibliography{ref.bib}% common bib file
\end{document}